\documentclass[aps,preprint,onecolumn,12pt,nofootinbib]{revtex4}

\providecommand{\U}[1]{\protect\rule{.1in}{.1in}}

\usepackage[utf8]{inputenc}
\usepackage{makecell, multirow, tabularx, rotating}
\usepackage{enumerate}
\usepackage{float}
\usepackage{lipsum}
\usepackage{soul} % Added from the second preamble for text highlighting/striking

\usepackage{amsmath, amsfonts, amssymb, mathrsfs, bm, bbm}
\usepackage{tensor}
\usepackage{bigints}
\usepackage{mathtools}

\usepackage{graphicx, epsf, subfigure}
\usepackage{xcolor} % Uniformly use xcolor (replaces 'color' to avoid conflicts)
\usepackage{tikz}
\usetikzlibrary{tikzmark}

\usepackage{hyperref}
\hypersetup{
    colorlinks=true,              
    breaklinks=true,              
    citecolor=blue,               
    linkcolor=[rgb]{0,0.5,0.9},   
    urlcolor=red,                 
    filecolor=green               
}
\newcommand{\ie}{\begin{equation}}
\newcommand{\fe}{\end{equation}}

\newcommand{\mincir}{\raise-3.truept\hbox{\rlap{\hbox{$\sim$}}\raise4.truept\hbox{$<$}\ }}
\newcommand{\magcir}{\raise-3.truept\hbox{\rlap{\hbox{$\sim$}}\raise4.truept\hbox{$>$}\ }}

\newif\ifshowchanges
\showchangestrue    % Uncomment this line = show track changes
\usepackage[normalem]{ulem} % [normalem] is extremely important to prevent overwriting italics
\usepackage{cancel}

\ifshowchanges
    \usepackage{changes}
    
    \newcommand{\redsout}[1]{{\color{gray}\bgroup\markoverwith{\textcolor{red}{\rule[0.5ex]{2pt}{0.4pt}}}\ULon{#1}}}
    
    \newcommand{\delmath}[1]{\textcolor{gray}{\cancel{#1}}} 
    
\else
    \usepackage[final]{changes}

    \newcommand{\redsout}[1]{}
    \newcommand{\delmath}[1]{} 
    
\fi

\definecolor{lime}{HTML}{A6CE39}
\DeclareRobustCommand{\orcidicon}{%
	\begin{tikzpicture}
	\draw[lime, fill=lime] (0,0) 
	circle [radius=0.16] 
	node[white] {{\fontfamily{qag}\selectfont \tiny ID}};
	\draw[white, fill=white] (-0.0625,0.095) 
	circle [radius=0.007];
	\end{tikzpicture}
	\hspace{-2mm}
}

\foreach \x in {A, ..., Z}{%
	\expandafter\xdef\csname orcid\x\endcsname{\noexpand\href{https://orcid.org/\csname orcidauthor\x\endcsname}{\noexpand\orcidicon}}
}

\begin{document}

\title{\Large{Optical Phenomena in a Non-Commutative Kalb-Ramond Black Hole Spacetime}}

%%%%%%%%%%%%%%%%%%%%%%%%%%%%%%%%%%%%%%%%%%%%%%%%%%%%%%%%%%%%%%%%%%%%%%%%%%%%%%%%%%%%%%%%%%%%%%%%%%%%%%%%%%%%%%%%%%%%%%%%%%%%%%%%%%%%%%%%%%%%%%%%%%%%%%%%%%%%%%%%%%%%%%%%%%%%%%%%%%%%%%%%%%%%%%%%%%%%%%%%

\author{A. A. Ara\'{u}jo Filho\orcidB{}}
\email{dilto@fisica.ufc.br (The corresponding author)}

\affiliation{Departamento de Física, Universidade Federal da Paraíba, Caixa Postal 5008, 58051-970, João Pessoa, Paraíba, Brazil}
\affiliation{Departamento de Física, Universidade Federal de Campina Grande Caixa Postal 10071, 58429-900 Campina Grande, Paraíba, Brazil.}

%%%%%%%%%%%%%%%%%%%%%%%%%%%%%%%%%%%%%%%%%%%%%%%%%%%%%%%%%%%%%%%%%%%%%%%%%%%%%%%%%%%%%%%%%%%%%%%%%%%%%%%%%%%%%%%%%%%%%%%%%%%%%%%%%%%%%%%%%%%%%%%%%%%%%%%%%%%%%%%%%%%%%%%%%%%%%%%%%%%%%%%%%%%%%%%%%%%%%%%%%%%%%%%%%%%%%%%%%%%%%%%%%%%%%%%%%%%%%%%%
\author{N. Heidari\orcidA{}}
\email{heidari.n@gmail.com}

\affiliation{Center for Theoretical Physics, Khazar University, 41 Mehseti Street, Baku, AZ-1096, Azerbaijan.}
\affiliation{Departamento de Física, Universidade Federal de Campina Grande Caixa Postal 10071, 58429-900 Campina Grande, Paraíba, Brazil.}
\affiliation{School of Physics, Damghan University, Damghan, 3671641167, Iran.}

%%%%%%%%%%%%%%%%%%%%%%%%%%%%%%%%%%%%%%%%%%%%%%%%%%%%%%%%%%%%%%%%%%%%%%%%%%%%%%%%%%%%%%%%%%%%%%%%%%%%%%%%%%%%%%%%%%%%%%%%%%%%%%%%%%%%%%%%%%%%%%%%%%%%%%%%%%%%%%%%%%%%%%%%%%%%%%%%%%%%%%%%%%%%%%%%%%%%%%%%%%%%%%%%%%%%%%%%%%%%%%%%%%%%%%%%%%%%%%%%%%%%%%%%%%%%%%%%%%%%%%%%%%%%%%%%%%%%%%%%

\author{Iarley P. Lobo\orcidD{}}
\email{lobofisica@gmail.com}

\affiliation{Department of Chemistry and Physics, Federal University of Para\'iba, Rodovia BR 079 - km 12, 58397-000 Areia-PB,  Brazil.}
\affiliation{Departamento de Física, Universidade Federal de Campina Grande Caixa Postal 10071, 58429-900 Campina Grande, Paraíba, Brazil.}

%%%%%%%%%%%%%%%%%%%%%%%%%%%%%%%%%%%%%%%%%%%%%%%%%%%%%%%%%%%%%%%%%%%%%%%%%%%%%%%%%%%%%%%%%%%%%%%%%%%%%%%%%%%%%%%%%%%%%%%%%%%%%%%%%%%%%%%%%%%%%%%%%%%%%%%%%%%%%%%%%%%%%%%%%%%%%%%%%%%%%%%%%%%%%%%%%%%%%%%%%%%%%%%%%%%%%%%%%%%%%%%%%%%%%%%%%%%%%%%%%%%%%%%%%%%%%%%%%%%%%%%%%%%%%%%%%%%%%%%%

\author{Yuxuan Shi\orcidE{}}
\email{shiyx2280771974@gmail.com}
\affiliation{Department of Physics, East China University of Science and Technology, Shanghai 200237, China}

%%%%%%%%%%%%%%%%%%%%%%%%%%%%%%%%%%%%%%%%%%%%%%%%%%%%%%%%%%%%%%%%%%%%%%%%%%%%%%%%%%%%%%%%%%%%%%%%%%%%%%%%%%%%%%%%%%%%%%%%%%%%%%%%%%%%%%%%%%%%%%%%%%%%%%%%%%%%%%%%%%%%%%%%%%%%%%%%%%%%%%%%%%%%%%%%%%%%%%%%%%%%%%%%%%%%%%%%%%%%%%%%%%%%%%%%%%%%%%%%%%%%%%%%%%%%%%%%%%%%%%%%%%%%%%%%%%%%%%%%

\begin{abstract}

This work investigates additional gravitational features of a newly proposed black hole spacetime within Kalb--Ramond gravity, incorporating non--commutative corrections arising from a gauge--theoretic approach recently introduced in the literature [arXiv:2507.17390]. Accordingly, null geodesics are solved numerically to trace photon paths; the photon sphere and shadow are determined. From Event Horizon Telescope (EHT) measurements of $Sgr A^{*}$, constraints on the parameters $\Theta$ (which encapsulates the non--commutativity) and $\ell$ (the Lorentz--violating parameter) are established. To examine the stability of critical orbits and the deflection angle (gravitational lensing) in the weak field scenario, we compute the Gaussian curvature in order to use the Gauss--Bonnet theorem. Moreover, the deflection angle has been calculated as well in the strong deflection limit. Furthermore, Lensing observables are estimated using EHT data for $Sgr A^{*}$ and $M87$. Topological features such as the topological photon sphere are also explored.

\end{abstract}
\maketitle

\tableofcontents

%%%%%%%%%%%%%%%%%%%%%%%%%%%%%%%%%%%%%%%%%%%%%%%%%%%%%%%%%%%%%%%%%%%%%%%%%%%%%%%%%%%%%%%%%%%%%%%%%%%%%%%%%%%%%%%%%%%%%%%%%%%%%%%%%%%%%%%%%%%%%%%%%%%%%%%%%%%%%%%%%%%%%%%%%%%%%%%%%%%%%%%%%%%%%%%%%%%%%%%%%%%%%%%%%%%%%%%%%%%%%%%%%%%%%%%%%%%%%%%%%%%%%%%%%%%%%%%%%%%%%%%%%%%%%%%%%%%%%%%%%%%%%%%%%%%%%%%%%%%%%%%%%%%%%%%%%%%%%%%%%%%%%%%%%%%%%%%%%%%%%%%%%%%%%%%%%%%%%%%%%%%%%%%%%%%%%%%%%%%%%%%%%%%%%%%%%%%%%%%%%%%%%%%%%%%%%%%%

\pagebreak
    
\section{Introduction}

Although the well--known general relativity permits arbitrarily precise spatial measurements, several quantum gravity scenarios suggest that space cannot be probed beyond a certain minimal scale, generally associated with the Planck length. This perspective has given rise to the concept of the so--called non--commutative spacetime, where position coordinates fail to commute—a feature originally inspired by developments in string theory \cite{szabo2003quantum,3,szabo2006symmetry}. Beyond it, non--commutative structures have been thoroughly explored within supersymmetric Yang--Mills frameworks, which accounts for a notable scenario for addressing finiteness and renormalization issues \cite{ferrari2003finiteness}. When gravity is considered, non--commutative modifications are typically introduced through the Seiberg--Witten map, which systematically deforms the gauge structure of spacetime and alters its corresponding symmetries \cite{chamseddine2001deforming}.

The incorporation of non--commutative geometry into gravitational studies has opened up alternative approaches to exploring black hole configurations, challenging traditional spacetime descriptions and introducing quantum--inspired modifications \cite{mann2011cosmological,2,heidari2023gravitational,karimabadi2020non,nicolini2009noncommutative,araujo2025comment,zhao2023quasinormal,heidari2024exploring,1,lopez2006towards,modesto2010charged,campos2022quasinormal,Heidari:2025sku}. These geometrical deformations have been widely applied to examine how black hole properties evolve under such conditions, particularly regarding the emission of radiation and the end-point of evaporation \cite{Filho:2025fvl,23araujo2023thermodynamics,myung2007thermodynamics} and neutrino physics \cite{AraujoFilho:2024mvz,AraujoFilho:2025rzh}. Numerous theoretical models have been developed to evaluate these effects and assess their implications for black hole stability and quantum gravitational corrections \cite{sharif2011thermodynamics,nozari2007thermodynamics,nozari2006reissner,lopez2006towards,banerjee2008noncommutative}.

Non--commutative spacetime emerges from a reformulation of the basic coordinate structure, in which classical commutativity is replaced by the algebraic relation $[x^\mu, x^\nu] = \mathbbm{i} \Theta^{\mu \nu}$. Here, $\Theta^{\mu \nu}$ is an antisymmetric matrix that encapsulates the degree and nature of the deformation. This fundamental shift motivates alternative geometric constructions, leading to a variety of techniques for embedding their features into gravity. In this approach, the first formulation of the Schwarzschild geometry was presented by Ref. \cite{chaichian2008corrections} within a setting where spacetime gauge symmetry is deformed. The construction relies on extending the de Sitter group SO(4,1) and combining it with the Poincaré symmetry ISO(3,1), with consistency ensured by the use of the Seiberg--Witten map.

Furthermore, a different approach emerged for the sake of introducing non--commutativity within the framework of GR was developed in \cite{nicolini2006noncommutative}, which focused on altering the content of matter rather than modifying the geometric part of the Einstein field equations. Instead of modeling the gravitational source as a point-like mass approach, their procedure replaced it with a smeared energy distribution. In other words, they considered non--commutative features directly into the stress--energy tensor. This method led to a non--singular matter profile that spread the mass over a finite region of space. Fundamentally, two primary choices for the smeared mass density have been proposed: a Gaussian distribution, $\rho_\Theta = M (4\pi \Theta)^{-3/2} e^{-r^2/4\Theta}$, and a Lorentzian profile, $\rho_\Theta = M \sqrt{\Theta} \pi^{-3/2} (r^2 + \pi \Theta)^{-2}$.

Jurić et al. \cite{Juric:2025kjl} have recently advanced a revised formulation for black hole solutions within the context of non--commutative gauge theory of gravity. Accordingly, their study reanalyzed the earlier framework proposed by Chaichian et. al \cite{chaichian2008corrections}, bringing to light a significant omission in the original treatment of non--commutative effects. The authors demonstrated that a crucial correction term had been excluded from the initial model, which compromised the consistency of the solution. By incorporating this previously overlooked element, namely, 
\ie
 - \frac{1}{16} \Theta^{\nu \rho} \Theta^{\lambda\tau} \Big[ \tilde{\omega}^{ac}_{\nu} \Tilde{\omega}^{cd}_{\lambda} \Big(  D_{\tau}R^{d5}_{\rho\mu} + \partial_{\tau}R^{d5}_{\rho\mu}   \Big)   \Big],
\nonumber
\fe
they reconstructed, therefore, a more complete version of the theory.

In other words, the updated framework presented in \cite{Juric:2025kjl} revealed extra terms that substantially impacted the underlying tetrad configuration, leading to corresponding changes in the metric tensor components. Based on this new formulation available in the literature, a recent investigation proposed a new black hole solution inspired by Kalb--Ramond gravity, incorporating non--commutative effects through a particula Moyal--type deformation ($\partial_r \wedge \partial_\theta$) \cite{heidari2025non}.

Recent theoretical research has revealed two separate configurations of Kalb–Ramond black holes. The first, presented in \cite{Yang:2023wtu}, has been extensively explored from various physical perspectives. Studies have addressed quasinormal modes \cite{araujo2024exploring}, strong–field lensing effects \cite{junior2024gravitational}, and greybody spectra \cite{guo2024quasinormal}. This geometry has also been employed to investigate spontaneous symmetry–breaking scenarios through constraint analyses \cite{junior2024spontaneous}, kinetic theory applications involving Vlasov gas accretion \cite{jiang2024accretion}, and particle motion in circular orbits together with quasi–periodic oscillations \cite{jumaniyozov2024circular}. Additional developments include particle production mechanisms \cite{12araujo2024particle}, configurations with global monopoles \cite{belchior2025global}, and slowly rotating generalizations \cite{Liu:2024lve}. Extensions of this background have incorporated electric charge \cite{duan2024electrically} and examined thermodynamics, lensing properties, radiation profiles, and topological aspects \cite{aa2024antisymmetric,al-Badawi:2024pdx,hosseinifar2024shadows,Zahid:2024ohn,Pantig:2025eda,heidari2024impact,chen2024thermal}, as well as global monopole \cite{baruah2025quasinormal} and ModMax \cite{sekhmani2025kalb} versions. A second solution, introduced later in \cite{Liu:2024oas}, has been employed to study entanglement degradation in Lorentz–violating frameworks \cite{liu2024lorentz} and to analyze quantum processes such as particle emission and greybody factors \cite{12araujo2024particle,12araujo2025does}.

The detection of gravitational waves by the LIGO and Virgo collaborations \cite{016,017,018} marked a turning point in gravitational physics, enabling examinations of phenomena such as light bending in weak--field conditions \cite{020,019}. Initially, gravitational lensing was predominantly studied within a cosmological framework, with light deflection analyzed using the Schwarzschild metric as a reference model \cite{021}. This line of inquiry was later broadened to include a range of spherically symmetric, static solutions \cite{022}. However, in scenarios where gravity become strong enough — for instance, in the vicinity of heavy black holes— the deflection of light is greatly amplified. Under such conditions, the complexity of the spacetime geometry requires more refined tools, including both analytical approximations and high--precision numerical methods, to accurately takes into account the behavior of light trajectories.

The success in extracting a black hole’s shadow by the Event Horizon Telescope has brought renewed attention to the study of gravitational lensing and its relevance to observational astrophysics \cite{026,029,027,028,024,025,023}. This groundbreaking image has amplified the need for precise models capable of describing how light behaves in the extreme curvature near compact objects. A notable step in this direction was made by Virbhadra and Ellis, who developed a simplified method for analyzing lensing in spacetimes that are asymptotically flat \cite{031,030}. Their formulation demonstrated that strong gravitational fields can bend light in such a way that a single luminous source may be seen as multiple distinct images.

Extensive progress in understanding gravitational lensing under strong--field conditions has emerged from the development of advanced analytical frameworks by Fritelli et al. \cite{032}, Bozza and his collaborators \cite{034,033}, and more recently, by Tsukamoto \cite{035} and Igata \cite{igata2025deflection}. These formalisms were specifically designed to handle light propagation in regions dominated by intense spacetime curvature. Their techniques have been applied successfully across a variety of gravitational backgrounds, ranging from spherically symmetric spacetimes \cite{r1,r2,r3,r4,r5,r6,r7,r8,cunha2018shadows,metcalf2019strong,oguri2019strong,grespan2023strong,bisnovatyi2017gravitational,Pantig:2022ely,ezquiaga2021phase,Ovgun:2018tua,Koyuncu:2014nga,Donmez:2023wtf,virbhadra2002gravitational,Pantig:2022gih,Donmez:2024lfi} to rotating and axisymmetric metrics \cite{37.4,hsieh2021strong,37.3,37.2,37.1,37.5,37.6,hsieh2021gravitational}, as well as to non--standard configurations like wormholes \cite{38.1,ovgun2019exact,38.5,38.3,38.4,38.2}. These methods have also been instrumental in probing deviations from general relativity, allowing to explore light deflection in the context of various modified gravity scenarios \cite{nascimento2024gravitational,soares2025light,chakraborty2017strong,40,araujo2025antisymmetric,heidari2024absorption,soares2025light}, and in electrically charged settings described by Reissner--Nordström geometry \cite{036.2,036.1,tsukamoto2023gravitational,zhang2024strong,036}.

Therefore, in this work, the analysis begins with a numerical integration of the null geodesic equations, from which photon trajectories are obtained. This procedure allows for determining the photon sphere and, consequently, the shadow of the black hole. Observational measurements from the Event Horizon Telescope (EHT) are then employed—using data for both $Sgr A^{*}$ and $M87$—to place bounds on the Lorentz–violating parameter $\ell$ and the non–commutative parameter $\Theta$. Gravitational lensing is investigated in the weak--field method through the computation of the Gaussian curvature and the application of the Gauss--Bonnet theorem, while the strong deflection limit is examined separately. The study also deals with topological aspects, including the characterization of the photon sphere topology.

%%%%%%%%%%%%%%%%%%%%%%%%%%%%%%%%%%%%%%%%%%%%%%%%%%%%%%%%%%%%%%%%%%%%%%%%%%%%%%%%%%%%%%%%%%%%%%%%%%%%%%%%%%%%%%%%%%%%%%%%%%%%%%%%%%%%%%%%%%%%%%%%%%%%%%%%%%%%%%%%%%%%%%%%%%%%%%%%%%%%%%%%%%%%%%%%%%%%%%%%%%%%%%%%%%%%%%%%%%%%%%%%%%%%%%%%%%%%%%%%%%%%%%%%%%%%%%%%%%%%%%%%%%%%%%%%%%%%%%%%%%%%%%%%%%%%%%%%%%%%%%%%%%%%%%%%%%%%%%%%%%%%%%%%%%%%%%%%%%%%%%%%%%%%%%%%%%%%%%%%%%%%%%%%%%%%%%%%%%%%%%%%%%%%%%%%%%%%%%%%%%%%%%%%%%%%%%%%%%%%%%%%%%%%%%%%%%%%%%%%%%%%%%%%%%%%%%%%%%%%%%%%%%%%%%%%%%%%%%%%%%%%%%%%%%%%%%%%%%%%%%%%%%%%%%%%%%%%%%%%%%%%%%%%%%%%%%%%%%%%%%%%%%%%%%%%%%%%%%%%%%%%%%%%%%%%%%%%%%%%%%%%%%%%%%%%%%%%%%%%%%%%%%%%%%%%%%%%%%%%%%%%%%%%%%%%%%%%%%%%%%%%%%%%%%%%%%%%%%%%%%%%%%%%%%%%%%%%%%%%%%%%%%%%%%%%%%%%%%%%%%%%%%%%%%%%%%%%%%%%

\section{\label{Sec2}The non-commutative Kalb-Ramond spacetime }

A new black hole solution has recently emerged in the literature within the framework of Kalb--Ramond gravity, incorporating non--commutative corrections from a gauge theory. Its general form is given by: \cite{heidari2025non}
\ie
\label{metrictensorss}
\mathrm{d}s^{2} = g_{\mu\nu}\left(x,\Theta\right) \mathrm{d}x^{\mu} \mathrm{d}x^{\nu}   = - A(\Theta,\ell) \mathrm{d}t^{2} +  B(\Theta,\ell) \mathrm{d}r^{2} + C(\Theta,\ell) \mathrm{d}\theta^{2} + D(\Theta,\ell) \mathrm{d}\varphi^{2},
\fe
with
\ie 
A(\Theta,\ell) = \frac{1}{1-\ell} - \frac{2 M}{r} - \frac{\Theta ^2 M (11 (\ell-1) M+4 r)}{2 (\ell-1) r^4},
\fe
\ie 
B(\Theta,\ell) = \frac{1}{\frac{1}{1-\ell}-\frac{2 M}{r}} + \frac{\Theta ^2 (\ell-1) M (3 (\ell-1) M+2 r)}{2 r^2 (2 (\ell-1) M+r)^2},
\fe
\ie 
C(\Theta,\ell) = r^2 -\frac{\Theta ^2 \left(64 (\ell-1)^2 M^2+32 (\ell-1) M r+r^2\right)}{16 (\ell-1) r (2 (\ell-1) M+r)},
\fe
\ie 
D(\Theta,\ell) = r^2 \sin ^2(\theta ) +\frac{1}{16} \Theta ^2 \left[5 \cos ^2(\theta )+\frac{4 \sin ^2(\theta ) \left(-2 (\ell-1) M^2+4 (\ell-1) M r+r^2\right)}{r (2 (\ell-1) M+r)}\right].
\fe

Based on the above metric, we proceed to analyze several remaining features of this black hole. Our focus will be fundamentally on the behavior of light, including the study of null geodesics, photon spheres, black hole shadows, gravitational lensing, and the structure of the topological photon sphere. {In this model, the parameter $\Theta$ controls the non--commutative deformation of the spacetime. More precisely, it emerges from the Moyal twist $\partial_r \wedge \partial_\theta$ adopted in the non--commutative gauge construction, and quantifies the ``strength'' of the corrections induced by the non--commuting coordinates. On the other hand, the parameter $\ell$ comes from the Kalb--Ramond gravitational sector and characterizes the Lorentz--violating contribution of the background field. In particular, $\ell$ is associated with the coupling between gravity and the vacuum expectation value of the antisymmetric Kalb--Ramond tensor, commonly written as $\ell=\xi\, b^{\mu\nu}b_{\mu\nu}/2$, where $\xi$ is the coupling constant and $b^{\mu\nu}$ denotes the background vacuum expectation value of the Kalb--Ramond field responsible for the spontaneous breaking of Lorentz symmetry.}

%%%%%%%%%%%%%%%%%%%%%%%%%%%%%%%%%%%%%%%%%%%%%%%%%%%%%%%%%%%%%%%%%%%%%%%%%%%%%%%%%%%%%%%%%%%%%%%%%%%%%%%%%%%%%%%%%%%%%%%%%%%%%%%%%%%%%%%%%%%%%%%%%%%%%%%%%%%%%%%%%%%%%%%%%%%%%%%%%%%%%%%%%%%%%%%%%%%%%%%%%%%%%%%%%%%%%%%%%%%%%%%%%%%%%%%%%%%%%%%%%%%%%%%%%%%%%%%%%%%%%%%%%%%%%%%%%%%%%%%%%%%%%%%%%%%%%%%%%%%%%%%%%%%%%%%%%%%%%%%%%%%%%%%%%%%%%%%%%%%%%%%%%%%%%%%%%%%%%%%%%%%%%%%%%%%%%%%%%%%%%%%%%%%%%%%%%%%%%%%%%%%%%%%%%%%%%%%%%%%%%%%%%%%%%%%%%%%%%%%%%%%%%%%%%%%%%%%%%%%%%%%%%%%%%%%%%%%%%%%%%%%%

\section{\label{Sec8}Light Propagation}

\subsection{Geodesics}

In this section we focus on analyzing the geodesic behavior of test particles in the given spacetime. As previously noted, computing the Christoffel symbols is a fundamental step in this process. Accordingly, we start by formulating:
\ie
\frac{\mathrm{d}^{2}x^{\mu}}{\mathrm{d}\tau^{2}} + \Gamma\indices{^\mu_\alpha_\beta}\frac{\mathrm{d}x^{\alpha}}{\mathrm{d}\tau}\frac{\mathrm{d}x^{\beta}}{\mathrm{d}\tau} = 0. \label{geogeo}
\fe

In our notation, $\tau$ denotes an arbitrary affine parameter governing the evolution along the particle's worldline. The resulting formulation gives rise to a set of four interdependent differential equations, each corresponding to motion along one of the spacetime coordinates, and can be written as follows:
\ie
\frac{\mathrm{d} t^{\prime}}{\mathrm{d} \tau} = \frac{4 M r' t' \left(11 \Theta ^2 (\ell-1) M+(\ell-1) r^3+3 \Theta ^2 r\right)}{r \left(11 \Theta ^2 (\ell-1) M^2+4 M r \left(\Theta ^2+(\ell-1) r^2\right)+2 r^4\right)},
\label{7}
\fe
\ie
\begin{split}
& \frac{\mathrm{d} r^{\prime}}{\mathrm{d} \tau} = \frac{1}{8 (\ell-1)^2 r^3 (2 (\ell-1) M+r) \left(-3 \Theta ^2 (\ell-1) M^2+4 (\ell-1) M r^3-2 \Theta ^2 M r+2 r^4\right)} \times \Bigg\{ \\
&  -8 (\ell-1)^2 M r^2 \Upsilon  \left(r'\right)^2  +16 M \left(t'\right)^2 (2 (\ell-1) M+r)^3 \left(11 \Theta ^2 (\ell-1) M+(\ell-1) r^3+3 \Theta ^2 r\right)  \\
& +(\ell-1) \nu  r^3 \left(\theta '\right)^2 (2 (\ell-1) M+r)+4 (\ell-1){\xi}  r^3 \sin ^2(\theta ) \left(\varphi '\right)^2 (2 (\ell-1) M+r) \Bigg\},
\end{split}
\fe
\ie
\begin{split}
& \frac{\mathrm{d} \theta^{\prime}}{\mathrm{d} \tau} =  \frac{(\ell-1)\sin (2 \theta ) \left(\varphi '\right)^2 \left(-8 \Theta ^2 (\ell-1) M^2+2 (\ell-1) M r \left(3 \Theta ^2+16 r^2\right)+16 r^4-\Theta ^2 r^2\right)}{-128 \Theta ^2 (\ell-1)^2 M^2+64 (\ell-1) M r \left((\ell-1) r^2-\Theta ^2\right)+32 (\ell-1) r^4-2 \Theta ^2 r^2} \\
& -\frac{2 \theta ' r' (\ell-1) \left(64 \Theta ^2 (\ell-1)^2 M^3+64 (\ell-1) M^2 r \left(\Theta ^2+(\ell-1) r^2\right)+M \left(64 (\ell-1) r^4+15 \Theta ^2 r^2\right)+16 r^5\right)}{r (2 (\ell-1) M+r) \left(-64 \Theta ^2 (\ell-1)^2 M^2+32 (\ell-1) M r \left((\ell-1) r^2-\Theta ^2\right)+16 (\ell-1) r^4-\Theta ^2 r^2\right)},
\end{split}
\fe
and 
\ie
\begin{split}
& \frac{\mathrm{d} \varphi^{\prime}}{\mathrm{d} \tau}  = \frac{2 \sin (\theta ) \varphi'}{4 \sin ^2(\theta ) \chi +5 \Theta ^2 r \cos ^2(\theta ) (2 (\ell-1) M+r)} \times \Bigg\{  \\
&    -\frac{4 \sin (\theta ) r' \left(2 \Theta ^2 (\ell-1)^2 M^3+2 (\ell-1) M^2 r \left(\Theta ^2+8 (\ell-1) r^2\right)+(\ell-1) M r^2 \left(16 r^2-\Theta ^2\right)+4 r^5\right)}{r (2 (\ell-1) M+r)} \\
& - \theta ' \cos (\theta ) \left(-8 \Theta ^2 (\ell-1) M^2+2 (\ell-1) M r \left(3 \Theta ^2+16 r^2\right)+16 r^4-\Theta ^2 r^2\right)\Bigg\},
\label{10}
\end{split}
\fe
where 
\ie
\chi \equiv \, -2 \Theta ^2 (\ell-1) M^2+4 (\ell-1) M r \left(\Theta ^2+2 r^2\right)+r^2 \left(\Theta ^2+4 r^2\right),
\fe
\ie
\Upsilon \equiv \, 6 \Theta ^2 (\ell-1)^2 M^2+4 (\ell-1) M r \left(2 \Theta ^2+(\ell-1) r^2\right)+2 (\ell-1) r^4+3 \Theta ^2 r^2,
\fe
\ie
\nu \equiv \, 64 \Theta ^2 (\ell-1)^2 M^3+64 (\ell-1) M^2 r \left(\Theta ^2+(\ell-1) r^2\right)+64 (\ell-1) M r^4+15 \Theta ^2 M r^2+16 r^5,
\fe
\ie
\xi \equiv \, 2 \Theta ^2 (\ell-1)^2 M^3+2 (\ell-1) M^2 r \left(\Theta ^2+8 (\ell-1) r^2\right)+(\ell-1) M r^2 \left(16 r^2-\Theta ^2\right)+4 r^5,
\fe
{where the prime symobl denotes differentiation with respect to the affine parameter $\tau$, i.e.
$x'^{\mu} \equiv \mathrm{d}x^{\mu}/\mathrm{d}\tau$.
Accordingly, $\mathrm{d}x'^{\mu}/\mathrm{d}\tau = \mathrm{d}^{2}x^{\mu}/\mathrm{d}\tau^{2}$, and Eqs.~(\ref{7})--(\ref{10}) are the explicit second-order geodesic equations written componentwise.}

\begin{figure}
    \centering
    \includegraphics[scale=0.51]{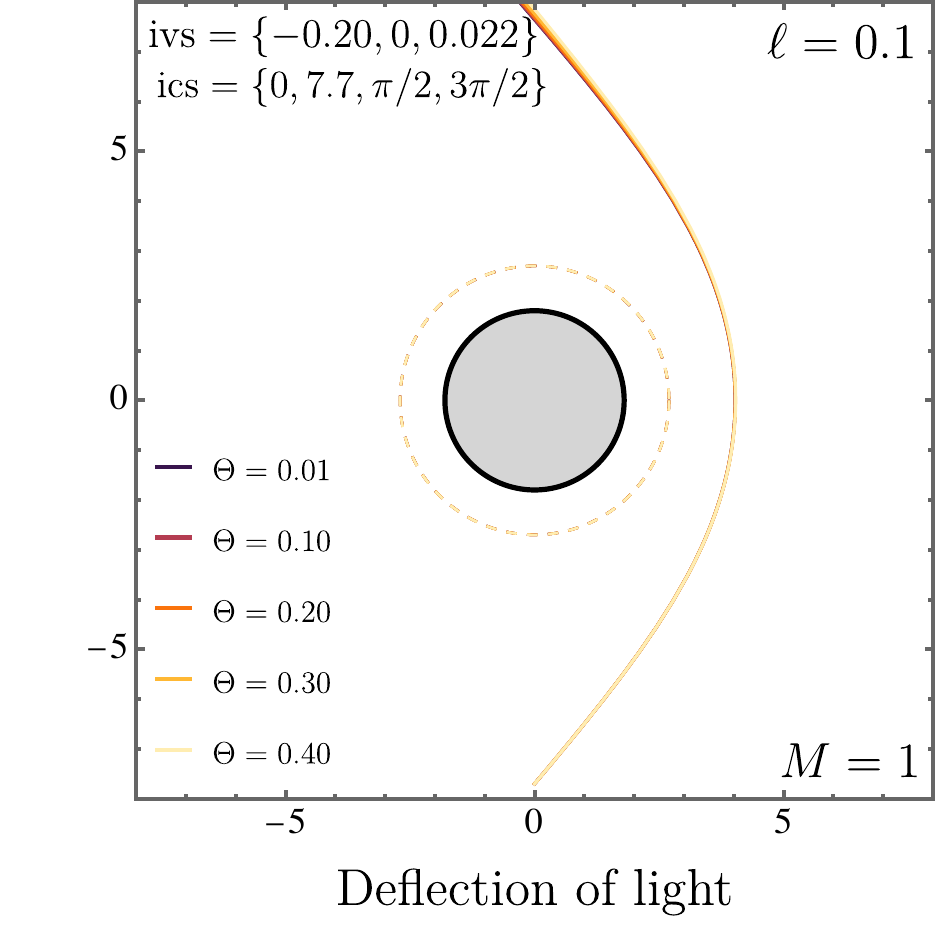}
    \includegraphics[scale=0.51]{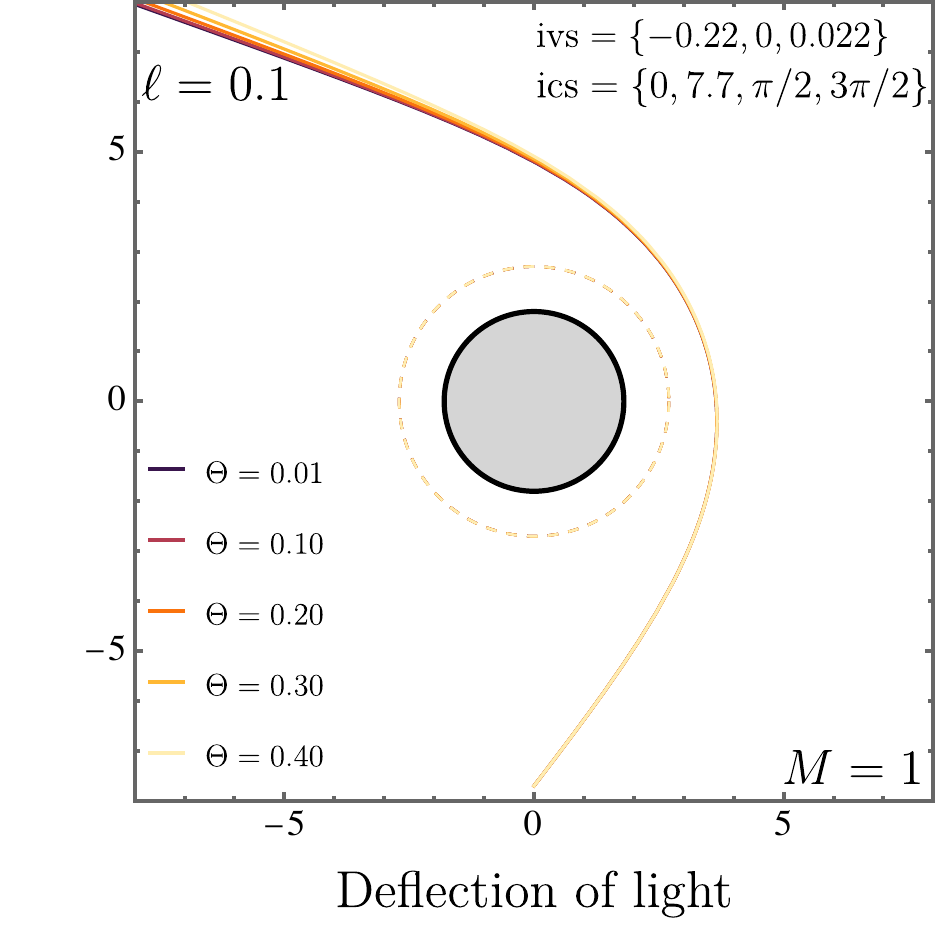}
    \includegraphics[scale=0.51]{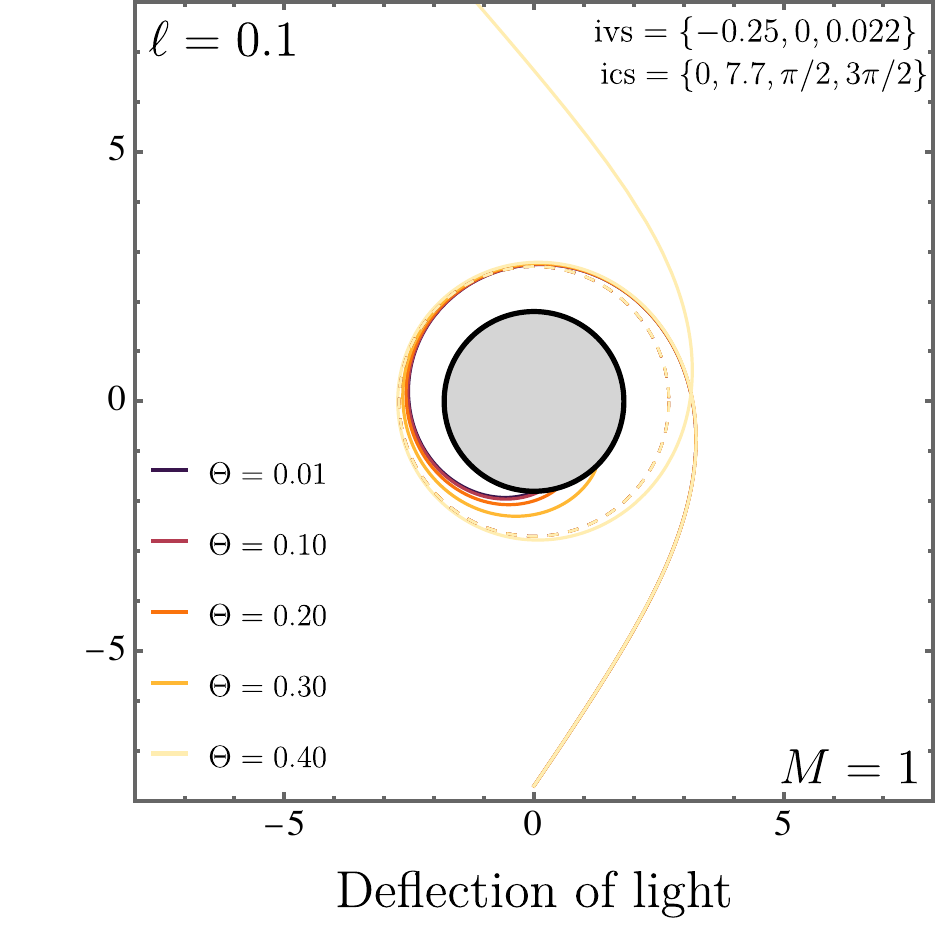}
    \includegraphics[scale=0.51]{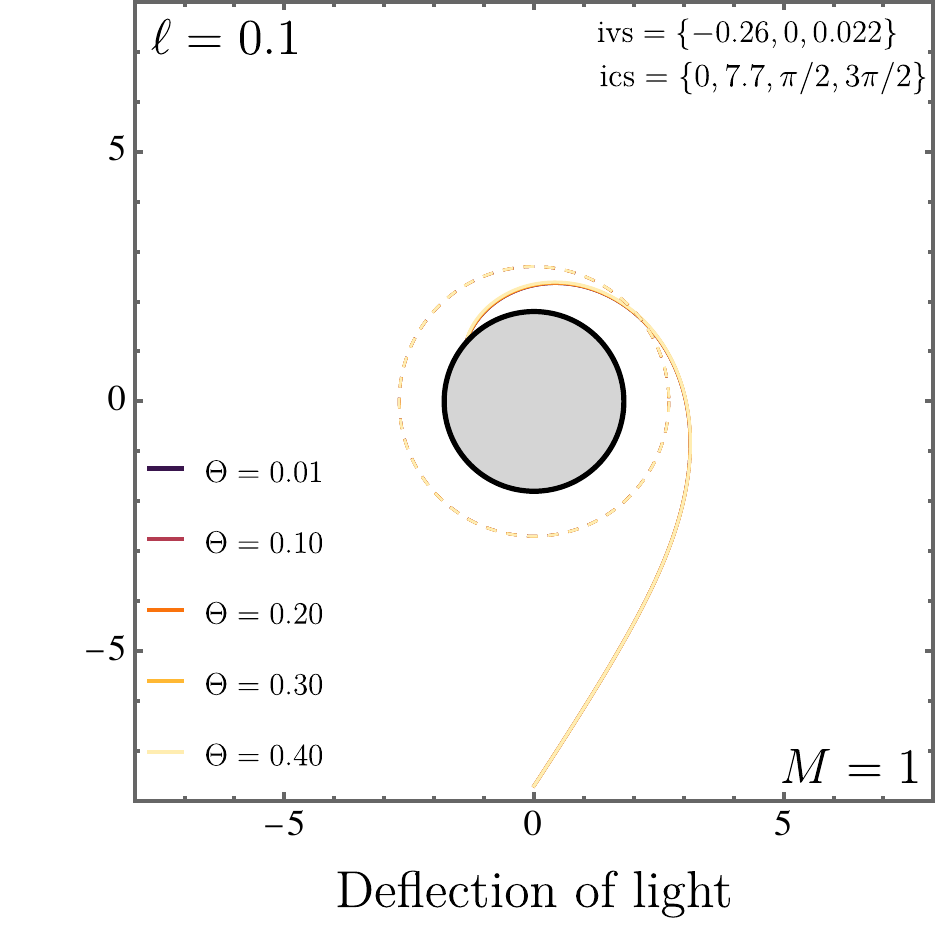}
    \caption{  {Light trajectories for increasing values of the non--commutative parameter $\Theta$, with $\ell=0.1$ and $M=1$ fixed. The black circular line marks the event horizon, the dashed curves denote the outermost unstable photon orbits, and the gray disk represents the black hole interior.} }
    \label{geooooo}
\end{figure}

Fig. \ref{geooooo} depicts how null geodesics behave for various values of the non--commutative parameter $\Theta$, with the mass fixed at $M = 1$ and the Lorentz--violating parameter set to $\ell = 0.1$. The parameter $\Theta$ ranges from $0.01$ to $0.4$ in incremental steps. Overall, increasing $\Theta$ leads to light rays following more ``open" paths, indicating weaker gravitational deflection. The simulations were carried out using a set of distinct numerical initial conditions, which are explicitly indicated within each plot. {The black circular line indicates the event horizon, while the dashed curves denote the outermost unstable\footnote{The stability of the critical orbits will be examined in detail in Sec.~\ref{Sec10}, devoted to gravitational lensing phenomena.} photon orbits. The gray disk at the center represents the black hole interior.}

\begin{figure}
    \centering
    \includegraphics[scale=0.643]{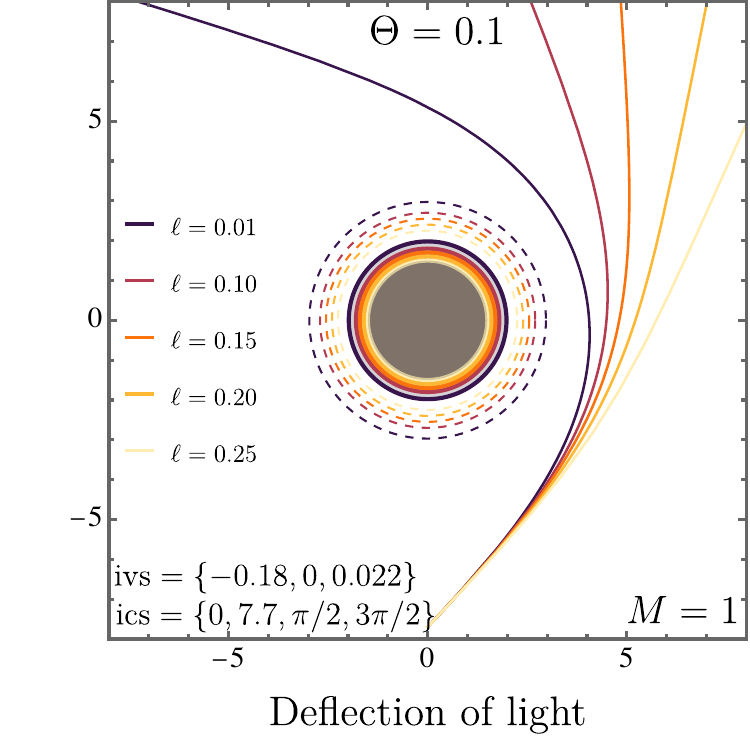}
    \includegraphics[scale=0.643]{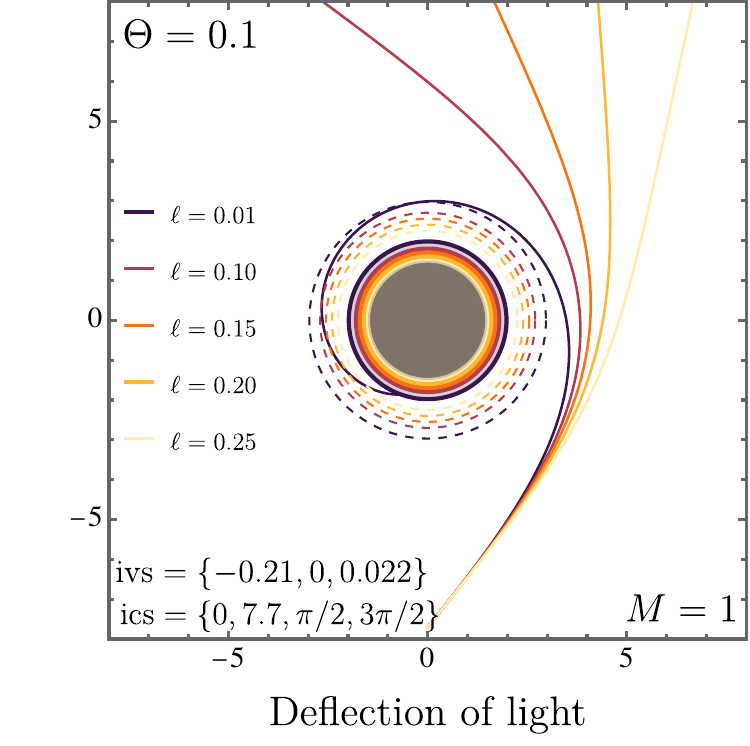}
    \includegraphics[scale=0.643]{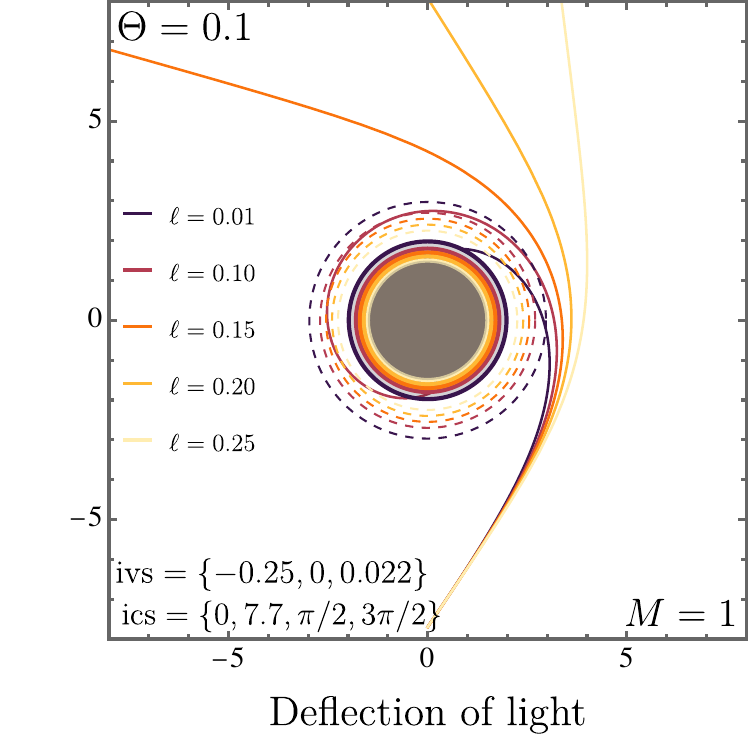}
    \includegraphics[scale=0.643]{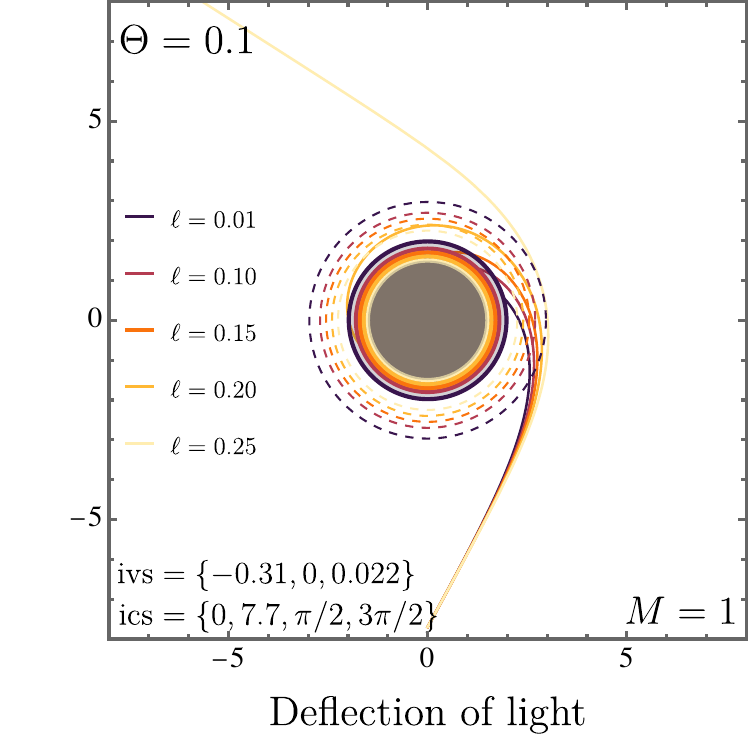}
    \caption{ {The light trajectories are shown for fixed values of the non--commutative parameter $\Theta$, while $\ell$ is varied and $M=1$. Different initial conditions were adopted in the numerical integration. The dashed lines denote the photon radii, the solid circles in the central region represent the event horizons, and the curved colored lines correspond to the light rays. } }
    \label{geooooo2}
\end{figure}

{Similarly, Fig.~\ref{geooooo2} illustrates the effect of the Lorentz--violating parameter $\ell$ on the propagation of light, with the non--commutative parameter fixed at $\Theta = 0.1$ and $M=1$. Different initial conditions were used in the numerical integration to generate the trajectories. The dashed lines indicate the photon radii, the solid circles in the central region represent the event horizons, and the curved colored lines correspond to the light rays. In agreement with the previous analysis, increasing $\ell$ leads to broader trajectories, indicating a behavior qualitatively similar to that produced by variations in $\Theta$.}

%%%%%%%%%%%%%%%%%%%%%%%%%%%%%%%%%%%%%%%%%%%%%%%%%%%%%%%%%%%%%%%%%%%%%%%%%%%%%%%%%%%%%%%%%%%%%%%%%%%%%%%%%%%%%%%%%%%%%%%%%%%%%%%%%%%%%%%%%%%%%%%%%%%%%%%%%%%%%%%%%%%%%%%%%%%%%%%%%%%%%%%%%%%%%%%%%%%%%%%%%%%%%%%%%%%%%%%%%%%%%%%%%%%%%%%%%%%%%%%%%%%%%%%%%%%%%%%%%%%%%%%%%%%%%%%%%%%%%%%%%%%%%%%%%%%%%%%%%%%%%%%%%%%%%%%%%%%%%%%%%%%%%%%%%%%%%%%%%%%%%%%%%%%%%%%%%%%%%%%%%%%%%%%%%%%%%%%%%%%%%%%%%%%%%%%%%%%%%%%%%%%%%%%%%%%%%%%%%%%%%%%%%%%%%%%%%%%%%%%%%%%%%%%%%%%%%%%%%%%%%%%%%%%%%%%%%%%%%%%%%%%%

\subsection{Photon sphere and black hole shadows}\label{sec:null}

The investigation of black hole shadows has risen as a central topic in contemporary gravitational physics \cite{zeng2022shadows, HAMIL2023101293, anacleto2023absorption, yan2023shadows}. This area has gained considerable momentum, particularly following the landmark imaging of the shadows of $Sgr A^{*}$ and $M87$ by the Event Horizon Telescope (EHT) \cite{ball2019first, gralla2021can, akiyama2019first}.

{For starting off, we begin with the spacetime geometry described in Eq.~(\ref{metrictensorss}). The dynamics of photon trajectories are then investigated using the Lagrangian formalism, formulated as
\ie
\mathcal{L} = \frac{1}{2}{g^{(\Theta,\ell)}_{\mu \nu }}{{\dot x}^\mu }{{\dot x}^\nu }.
\fe
In other words, we have
\begin{equation}
\label{lagrangian}
\mathcal{L} = \frac{1}{2}\Big[ - A(\Theta,\ell){{\dot t}^2} + B(\Theta,\ell){{\dot r}^2} + C(\Theta,\ell){{\dot \theta }^2} + D(\Theta,\ell){{\mathop{\rm \sin}\nolimits} ^2}\, \theta {{\dot \varphi }^2}\Big].
\end{equation}
By applying the Euler--Lagrange formalism and confining the analysis to the equatorial plane ($\theta = \frac{\pi}{2}$), the system yields two conserved quantities: the energy $E$ and the angular momentum $L$, as one should expect. These integrals of motion are given by the following expressions:
\begin{equation}\label{constant}
E = A(\Theta,\ell)\dot t \quad\mathrm{and}\quad L = D(\Theta,\ell)\dot \varphi,
\end{equation}
and regarding the massless modes, we have
\begin{equation}\label{light}
- A(\Theta,\ell){{\dot t}^2} + B(\Theta,\ell){{\dot r}^2} + D(\Theta,\ell){{\dot \varphi }^2} = 0.
\end{equation}
After carrying out the necessary algebraic steps to insert the expressions from Eq.(\ref{constant}) into Eq.(\ref{light}), the resulting form becomes:
\begin{equation}\label{rdot}
\frac{{{{\dot r}^2}}}{{{{\dot \varphi }^2}}} = {\left(\frac{{\mathrm{d}r}}{{\mathrm{d}\varphi }}\right)^2} = \frac{{D(\Theta,\ell)}}{{B(\Theta,\ell)}}\left(\frac{{D(\Theta,\ell)}}{{A(\Theta,\ell)}}\frac{{{E^2}}}{{{L^2}}} - 1\right).
\end{equation}
Furthermore, notice that
\ie
\frac{\mathrm{d}r}{\mathrm{d}\lambda} = \frac{\mathrm{d}r}{\mathrm{d}\varphi} \frac{\mathrm{d}\varphi}{\mathrm{d}\lambda}  = \frac{\mathrm{d}r}{\mathrm{d}\varphi}\frac{L}{D(\Theta,\ell)}, 
\fe
where
\ie
\Dot{r}^{2} = \left( \frac{\mathrm{d}r}{\mathrm{d}\lambda} \right)^{2} =\left( \frac{\mathrm{d}r}{\mathrm{d}\varphi} \right)^{2} \frac{L^{2}}{(D(\Theta,\ell))^{2}}
\fe
so that the effective potential $\mathrm{V}(\Theta,\ell)$ can properly be written below as
\ie
\label{potential}
\mathrm{V}(\Theta,\ell) = \frac{{D(\Theta,\ell)}}{{B(\Theta,\ell)}}\left(\frac{{D(\Theta,\ell)}}{{A(\Theta,\ell)}}\frac{{{E^2}}}{{{L^2}}} - 1\right)\frac{L^{2}}{\left( D(\Theta,\ell)\right)^{2}}.
\fe

With all these preliminaries established so far, let us determine the photon spheres. To do so, we have to impose the following condition:
\begin{equation}
\mathrm{V}(\Theta,\ell)=0, \quad\quad \frac{\mathrm{d} \,{\mathrm{V}(\Theta,\ell)}}{\mathrm{d}r} = 0 .
\end{equation}

By introducing the impact parameter $b_c=\frac{L}{E}$, the first condition leads to 
\ie
b_c=\frac{D(\Theta,\ell)}{A(\Theta,\ell)}\Big|_{r=r_{ph}}
\fe
Considering the above expression, the next condition leads the following expression

\begin{align}\nonumber
&8 r_{ph}^6 (r_{ph}+2 \ell -2)^2 (r_{ph}+3 \ell -3)+8 \Theta ^2 r_{ph}^3 (r_{ph}+2 \ell -2) \left(5 r_{ph}^2+r_{ph} (\ell -1) (\ell +26)\right.\\ \nonumber
&\left.+33 (\ell -1)^2\right)+\Theta ^4 \left(6 r_{ph}^4+54 r_{ph}^3 (\ell -1)+r_{ph}^2 (\ell -1) (169 \ell -173)\right.\\
&\left.+2 r_{ph} (\ell -1)^2 (88 \ell -107)-66 (\ell -1)^3\right)=0.
\end{align}

}

{The next step involves solving the above equation. Upon doing so, a single real and positive solution emerges, corresponding to the photon sphere radius, denoted by $r_{ph}$. Explicitly, this yields:}
\ie
r_{ph} \approx 3 M (1-\ell) + \frac{ \ell}{9 M}\Theta ^2,
\fe
{up to the second order of $\Theta$ and the first order of $\ell$.} {It is worth pointing out that the leading term on the right--hand side exactly matches the photon sphere radius obtained for the Kalb--Ramond black hole, as recently discussed in Refs.~\cite{araujo2024exploring,araujo2025particlasdasde,araujo2025does}.}

% \begin{table}[!ht]
%     \centering
%     \caption{{The radius of the outermost photon sphere is computed for different pairs of the parameters $\Theta$ and $\ell$, while keeping the black hole mass fixed at $M = 1$. }}
%     \begin{tabular}{|c|c|c|c|c|c|}
%     \hline
%          $r_{ph}$ & $\ell = 0.02$ & $\ell = 0.04$ & $\ell = 0.06$ & $\ell = 0.08$& $\ell = 0.10
%         $\\ \hline\hline
%         $\Theta = 0.0$ & 2.94000 & 2.88000 & 2.82000 & 2.76000 & 2.70000 \\ \hline
%         $\Theta= 0.2$ & 3.10956 & 3.04578 & 2.98200 & 2.91822 & 2.85444 \\ \hline
%         $\Theta = 0.4$& 3.61822 & 3.54311 & 3.46800 & 3.39289 & 3.31778 \\ \hline
%         $\Theta = 0.6$ & 4.46600 & 4.37200 & 4.27800 & 4.18400 & 4.09000\\ \hline
%         $\Theta = 0.8$ & 5.65289 & 5.53244 & 5.41200 & 5.29156 & 5.17111 \\ \hline
%         $\Theta = 0.99$ & 7.09454 & 6.94197 & 6.78940 & 6.63684 & 6.48428\\ \hline
%     \end{tabular}
%     \label{Tab:rphoton}
% \end{table}
%%%%%%%%%%%%%%%%%%%%
\begin{table}[!ht]
   \centering
    \caption{{The outermost photon sphere radius is computed for different pairs of the parameters $\ell$ and $\Theta$, while fixing the black hole mass as $M = 1$. }} \begin{tabular}{|c|c|c|c|c|c|}
    \hline
         $r_{ph}$ & $\ell = 0.02$ & $\ell = 0.04$ & $\ell = 0.06$ & $\ell = 0.08$& $\ell = 0.10
        $\\ \hline\hline
        $\Theta = 0.0$ & 2.9400 & 2.8800 & 2.8200 & 2.7600 & 2.7000 \\ \hline
        $\Theta= 0.2$ & 2.9401 & 2.8802 & 2.8203 & 2.7604 & 2.7005 \\ \hline
        $\Theta = 0.4$& 2.9403 & 2.8807 & 2.8211 & 2.7616 & 2.7021 \\ \hline
        $\Theta = 0.6$ & 2.9406 & 2.8814 & 2.8223 & 2.7633 & 2.7043\\ \hline
        $\Theta = 0.8$ & 2.9407 & 2.8821 & 2.8236 & 2.7653 & 2.7070 \\ \hline
    \end{tabular}
    \label{Tab:rphoton}
\end{table}
%%%%%%%%%%%%%%%%%%%%%%%%%%%%%%%%%

{Table \ref{Tab:rphoton} presents the numerical results for the radius of the photon sphere $r_{ph}$ of a black hole with mass fixed at $M = 1$, evaluated for various choices of the Lorentz--violating parameter $\ell$ and the non--commutative parameter $\Theta$. The data indicate that, when $\Theta$ is kept constant, larger values of $\ell$ cause $r_{ph}$ to decrease. In contrast, for a fixed $\ell$, increasing $\Theta$ results in a larger photon sphere radius.}

In addition, based on the procedures established in Refs .~\cite {perlick2015influence, konoplya2019shadow}, the angular radius of the black hole shadow in a spherically symmetric geometry can be cast as
\ie
\label{shadow}
\begin{split}
 & R_{sh}  =   \sqrt {\frac{{{D(\Theta,\ell)(r_{ph})}}}{{{A(\Theta,\ell)(r_{ph})}}}} = \sqrt{2} \sqrt{-\frac{(\ell-1) r_{ph}^4 \left(\frac{\Theta ^2 \left(-2 (\ell-1) M^2+4 (\ell-1) M r_{ph} + r_{ph}^2\right)}{4 r_{ph} (2 (\ell-1) M+r_{ph})}+r_{ph}^2\right)}{2 r_{ph}^3 (2 (\ell-1) M + r_{ph})+\Theta ^2 M (11 (\ell-1) M+4 r_{ph})}} \\
& =  \, \, \sqrt{\frac{\frac{\Theta ^6 \ell^2-81 \Theta ^2 (\ell-1) (3 \ell-1) M^4-18 \Theta ^4 (\ell-1) \ell M^2}{4 \Theta ^4 \ell^2+972 (\ell-1)^2 M^4-144 \Theta ^2 (\ell-1) \ell M^2}+\left(3 (\ell-1) M-\frac{\Theta ^2 \ell}{9 M}\right)^2}{\frac{18 M^2}{27 (\ell-1) M^2-\Theta ^2 \ell}+\frac{729 \Theta ^2 M^4 \left(9 (\ell-1) M^2-4 \Theta ^2 l\right)}{2 (\ell-1) \left(\Theta ^2 \ell-27 (\ell-1) M^2\right)^4}+\frac{1}{1-\ell}}}  \\
& \approx \,\,  3 \sqrt{3} M  -\frac{9}{2} \sqrt{3} M \ell - \frac{\Theta ^2}{8 \left(\sqrt{3} M\right)}   -\frac{\Theta ^2}{16 \left(\sqrt{3} M\right)}   \ell.
\end{split}
\fe
The expression above is expanded to first order in $\ell$ and to second order in $\Theta$, with the analysis restricted to the equatorial plane ($\theta = \pi/2$).

{There are two important observations worth emphasizing here. First, in the limit where both deformation parameters vanish, $\Theta, \ell \to 0$, the expression correctly reproduces the well-known Schwarzschild shadow radius, $3\sqrt{3}M$, as expected. Second, setting $\Theta \to 0$ while keeping $\ell \ne 0$ yields the shadow radius associated with the Kalb--Ramond black hole, given by $3\sqrt{3}M - \frac{9}{2}\sqrt{3} M \ell$, in agreement with previous findings \cite{araujo2025particlasdasde}.}

\begin{figure}[ht]
		\centering
		\includegraphics[width=99mm]{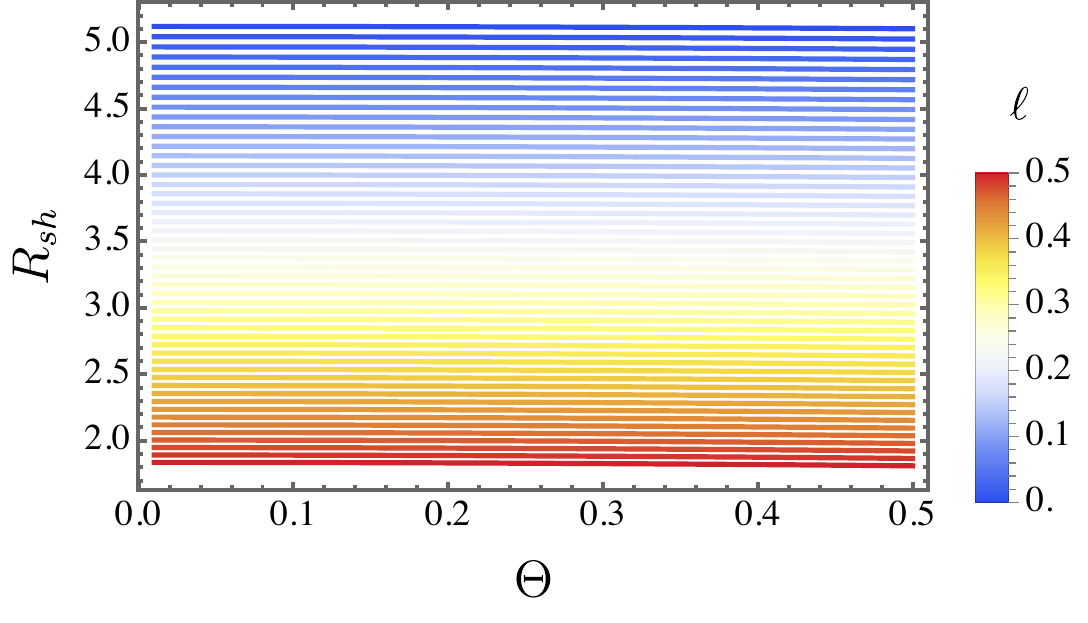}
        \includegraphics[width=99mm]{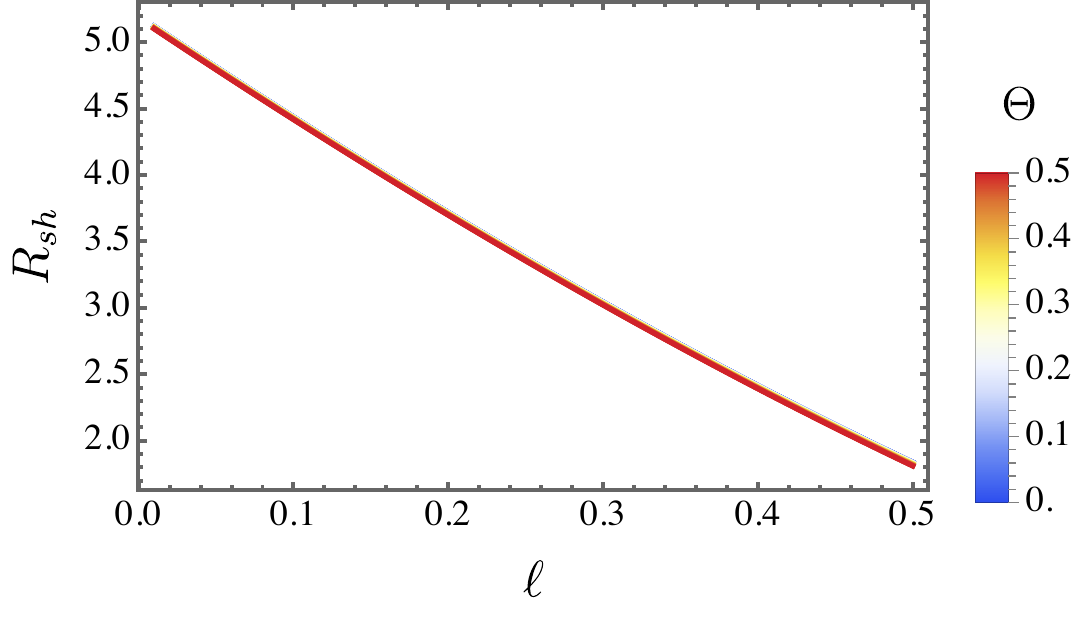}
		\caption{ The top panel illustrates the impact of the non--commutative parameter $\Theta$ on the shadow radius $R_{sh}$ for several fixed values of the Lorentz--violating parameter $\ell$. In contrast, the bottom panel explores the dependence of $R_{sh}$ on $\ell$ across different values of $\Theta$. Throughout both analyses, the black hole mass is held constant at $M = 1$.}
		\label{fig:Shadow}
	\end{figure}

{Regarding the evaluation of the shadow radius (both in the tabulated data and in the plots), the full expressions for $r_{ph}$ and consequently $R_{sh}$ have been used without employing any perturbative approximation to ensure accuracy and visual clarity. However, due to their considerable length, the exact analytical forms are not displayed. The corresponding numerical values are summarized in Tab.~\ref{tabshadows}.}
\begin{table}[!ht]
    \centering
    \caption{For a fixed black hole mass $M = 1$, the shadow radius $R_{sh}$ is evaluated for various combinations of the parameters $\Theta$ and $\ell$.}
    \begin{tabular}{|c|c|c|c|c|c|}
    \hline
         {$R_{sh}$} & ${\ell = 0.02}$ & ${\ell = 0.04}$ & ${\ell = 0.06}$ & ${\ell = 0.08}$& ${\ell = 0.10
        }$\\ \hline\hline
        $\Theta = 0.0$ & 5.04027 & 4.88438 & 4.72850 & 4.57261 & 4.41673 \\ \hline
        $\Theta= 0.2$ & 5.03735 & 4.88144 & 4.72553 & 4.56961 & 4.41370 \\ \hline
        $\Theta = 0.4$& 5.02861 & 4.87261 & 4.71661 & 4.56061 & 4.40461 \\ \hline
        $\Theta = 0.6$ & 5.01403 & 4.85788 & 4.70174 & 4.54559 & 4.38945\\ \hline
        $\Theta = 0.7$ & 5.00455 & 4.84831 & 4.69208 & 4.53584 & 4.37960 \\ \hline
    \end{tabular}
    \label{tabshadows}
\end{table}
To provide a visual interpretation, Fig.~\ref{fig:Shadow} presents the black hole shadow profiles for various combinations of the parameters $\Theta$ and $\ell$.  

The top panel illustrates how the shadow radius is impacted by the non--commutative parameter $\Theta$, for a set of fixed values of $\ell$, with the black hole mass held constant at $M = 1$. These results are consistent with the data provided in Tab~\ref{tabshadows}. One observes that, for increasing $\ell$, the shadow radius tends to shrink across the range of $\Theta$ considered.

Conversely, the bottom panel examines the dependence of the shadow radius on the Lorentz--violating parameter $\ell$, for various values of $\Theta$, again with $M = 1$ fixed. In this case, the trend reverses: the increase of $\Theta$ lead to an augment in the shadow radius for a given $\ell$.

Furthermore, the next section is devoted to constraining the parameters $\ell$ and $\Theta$ using observational measurements coming from EHT.

%%%%%%%%%%%%%%%%%%%%%%%%%%%%%%%%%%%%%%%%%%%%%%%%%%%%%%%%%%%%%%%%%%%%%%%%%%%%%%%%%%%%%%%%%%%%%%%%%%%%%%%%%%%%%%%%%%%%%%%%%%%%%%%%%%%%%%%%%%%%%%%%%%%%%%%%%%%%%%%%%%%%%%%%%%%%%%%%%%%%%%%%%%%%%%%%%%%%%%%%%%%%%%%%%%%%%%%%%%%%%%%%%%%%%%%%%%%%%%%%%%%%%%%%%%%%%%%%%%%%%%%%%%%%%%%%%%%%%%%%%%%%%%%%%%%%%%%%%%%%%%%%%%%%%%%%%%%%%%%%%%%%%%%%%%%%%%%%%

%%%%%%%%%%%%%%%%%%%%%%%%%%%
\subsection{Topological Structure of the Photon Sphere}

In recent years, topological approaches have become a robust and influential framework for analyzing the structure and stability of photon spheres~\cite{Wei2020,Cunha2020,Sadeghi2024,BahrozBrzo2025,alipour2024weak}. These approaches offer an elegant way to classify photon spheres as topological defects associated with vector fields constructed from metric potentials. In this section, we apply this method to the Kalb--Ramond black hole and explore the photon sphere and its associated topological charge.

{It is important to mention that the topological charge associated with the photon sphere should be understood as a geometric quantity that classifies the critical null orbits of the spacetime. Its physical meaning is tied to the role of the photon sphere as the boundary between capture and scattering trajectories: a small inward perturbation drives the photon into the black hole, whereas a small outward perturbation allows it to escape to infinity. In this sense, the topological charge does not represent a new conserved dynamical observable, but a global characterization of the structure of null geodesics near the compact object.}

Firstly, let us define the scalar potential function \( \mathcal{H}(r, \theta) \) that reflects the structure of null geodesics in a spherically symmetric background:
\begin{equation}
\mathcal{H}(r, \theta) = \sqrt{\frac{g_{\phi\phi}(r, \theta)}{-g_{tt}(r, \theta)}}= \sqrt{\frac{D{( \Theta,\ell)}}{A{( \Theta,\ell)}
}},
\label{eq:HR-KR}
\end{equation}
%%%%%%%%%%%%%%%%%%%%%%%%%%%%%%%%%%

%%%%%%%%%%%%%%%%%%%%%%%%%%%%%%%%%%%%%
%%%%%%%%%%%%%%%%%%%%%%%%%%%%
This potential reaches an extremum at the location of circular photon orbits, determined by the condition
\begin{equation}
\frac{\partial \mathcal{H}(r, \theta)}{\partial r} = 0.
\end{equation}

which leads to the following equation for a null geodesics in equatorial plane $\theta=\frac{\pi}{2}$.

\begin{align}\nonumber
   &8 {r_{ph}}^9 + 56 M (\ell - 1) {r_{ph}}^8 + 40 \Theta^2 M {r_{ph}}^6 + 6 \Theta^4 M {r_{ph}}^4 + 2 M^2 (\ell - 1) \Bigl[64 (\ell - 1) {r_{ph}}^7 \\  \nonumber
  &+4 \Theta^2 (\ell + 36) {r_{ph}}^5 + 27 \Theta^4 {r_{ph}}^3 \Bigr] + M^3 (\ell - 1) \Bigl[96 (\ell - 1)^2 {r_{ph}}^6 + 8 \Theta^2 (\ell - 1)(2 \ell + 85) {r_{ph}}^4  \\ \nonumber
  &+\Theta^4 (169 \ell - 173) {r_{ph}}^2\Bigr]+ 2 \Theta^2 M^4 (\ell - 1)^2 \Bigl[264 (\ell - 1) {r_{ph}}^3 + \Theta^2 (88 \ell - 107) r_{\text{ph}}\Bigr]\\
  & - 66 \Theta^4 M^5 (\ell - 1)^3=0.
\end{align}

The photon sphere is located at the critical radius $r = r_{ph}$, whose stability is determined by the curvature properties of $\mathcal{H}(r,\theta)$.

\begin{figure}[ht!]
    \centering
    \includegraphics[width=0.47\textwidth]{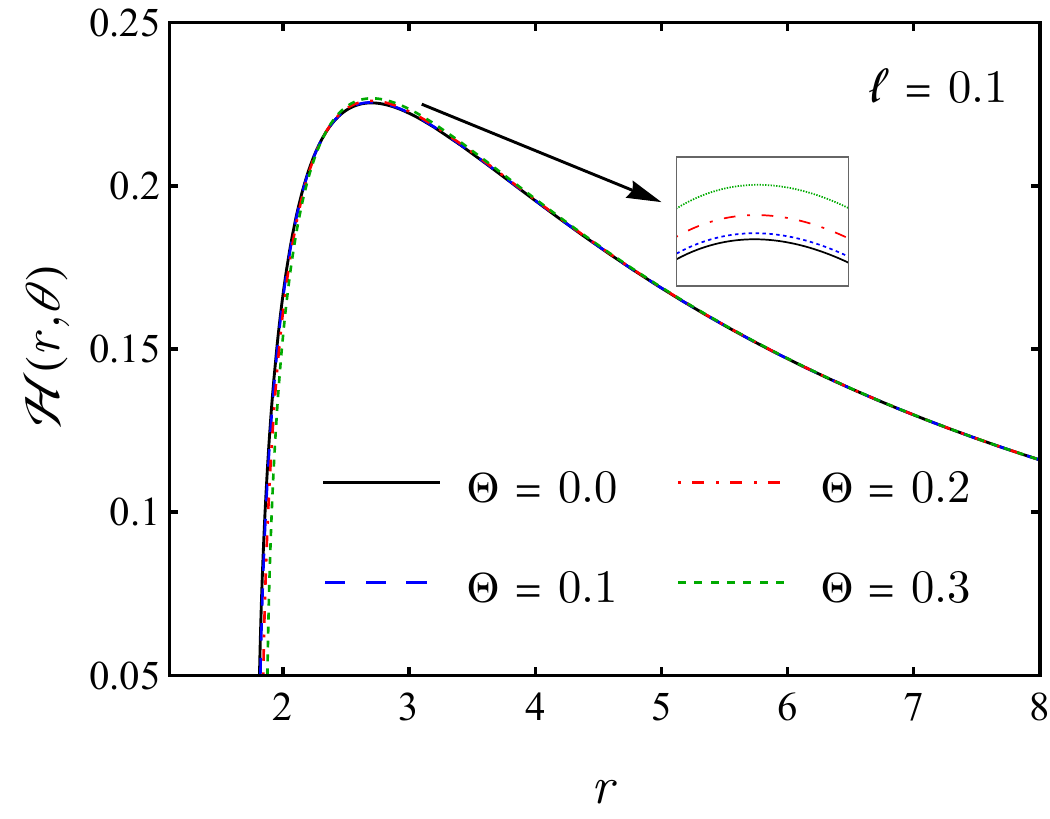}
    \includegraphics[width=0.47\textwidth]{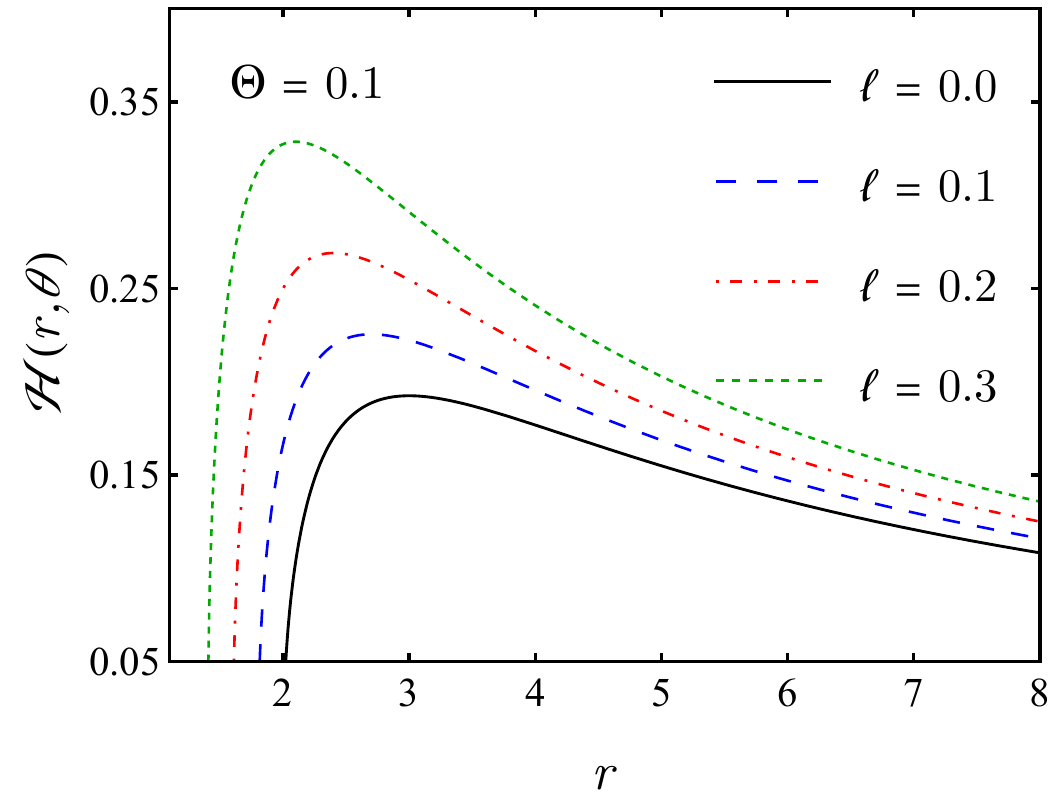}
    \caption{
    The topological potential $\mathcal{H}(r,\theta)$ with respect to the radial coordinate $r$ is shown. In the first panel, the Lorentz--violating parameter is set to $\ell = 0.1$ and the non--commutative parameter $\Theta$ takes different values. In the second panel, $\Theta$ is fixed at $0.1$ while $\ell$ is varied. In both cases, we consider $M = 1$.}
    \label{fig:KR_Hpotentials}
\end{figure}

%%%%%%%%%%%%%%%%%%%%%%%%%
In Fig.~\ref{fig:KR_Hpotentials}, we display the behavior of the topological potential function $\mathcal{H}(r,\theta)$. The plots clearly show a well-defined global maximum in $\mathcal{H}(r,\theta)$, located at the photon sphere radius. This maximum signifies an unstable circular orbit for photons.
The left panel shows that increasing the non--commutative parameter $\Theta$ slightly increase the peak height of the potential, and increases the maxumim place subtly. In contrast, the right panel reveals that the Lorentz--violating parameter $\ell$ has a more significant effect, reducing the photonic radius. This indicates that $\ell$ plays a more dominant role in modifying the $r_{ph}$ in this model.\\
%%%%%%%%%%%%%%%%%%%%%%%%%%%%%%%%%%%%%%%%%%%%

%%%%%%%%%%%%%%%%%%%%%%%%%%%%%%%%%%%%%%%%%%%%%%%%%%%%%%%%%%%%%%%%%%%%%%%%%%%%%%%%%%%%%%%%%%%%%%%%%%%

To proceed with the topological formulation, a two-dimensional vector field \( \vec{\varphi} = (\varphi^r, \varphi^\theta) \) is introduced as 
\begin{align}
\varphi^r &= \frac{1}{\sqrt{g_{rr}}} \, \partial_r \mathcal{H}(r, \theta), \label{eq:phi_r_KR} \\
\varphi^\theta &= \frac{1}{\sqrt{g_{\theta \theta}}} \, \partial_\theta \mathcal{H}(r, \theta). \label{eq:phi_theta_KR}
\end{align}
The normalized form of this field is expressed as
\begin{equation}
n^\alpha = \frac{\varphi^\alpha}{\|\vec{\varphi}\|}, \quad \alpha = r, \theta.
\end{equation}

Zero points of \( \vec{\varphi} \) correspond to locations where the photon sphere occurs, and the topological structure of these defects can be further analyzed through the notion of topological currents.

Following Duan's topological current theory~\cite{Duane1984}, we introduce the antisymmetric superpotential \( \Upsilon^{\mu\nu} \) defined in (2+1) dimensions as:
\begin{equation}
\Upsilon^{\mu\nu} = \frac{1}{2\pi} \epsilon^{\mu\nu\rho} \epsilon^{\alpha\beta} n_\alpha \partial_\rho n_\beta,
\end{equation}
here \( \alpha, \beta = 1,2 \) and \( \mu, \nu, \rho = 0,1,2 \). The associated topological current can be obtained via:
\begin{equation}
j^\mu = \partial_\nu \Upsilon^{\mu\nu} = \frac{1}{2\pi} \epsilon^{\mu\nu\rho} \epsilon^{\alpha\beta} \partial_\nu n_\alpha \partial_\rho n_\beta.
\end{equation}
This current is conserved (\( \partial_\mu j^\mu = 0 \)) and only at the zeros of the vector field, i.e.,\( \vec{\varphi} = 0 \), is non-zero. 

Using characteristics of the Dirac delta function and the Jacobian determinant \( J^\mu \), we may write:
\begin{equation}
j^\mu = J^\mu\left(\frac{\varphi}{x}\right)\delta^2(\varphi),
\end{equation}
where \( J^\mu \) encodes the Jacobian of the mapping \( \vec{\varphi} \colon (r, \theta) \to \mathbb{R}^2 \). The corresponding topological charge is then:
\begin{equation}
Q = \int_\Omega j^0\, \mathrm{d}^2x = \sum_i \omega_i,
\end{equation}
where each \( \omega_i \) is the winding number, is related to an isolated zero point of \( \vec{\varphi} \), computed via
\begin{equation}
\omega_i = \frac{1}{2\pi} \oint_{C_i} \mathrm{d}\Omega, \quad \Omega = \arg\left(\varphi^r + i\varphi^\theta\right).
\end{equation}

\begin{figure}[ht!]
    \centering
    \includegraphics[width=0.6\textwidth]{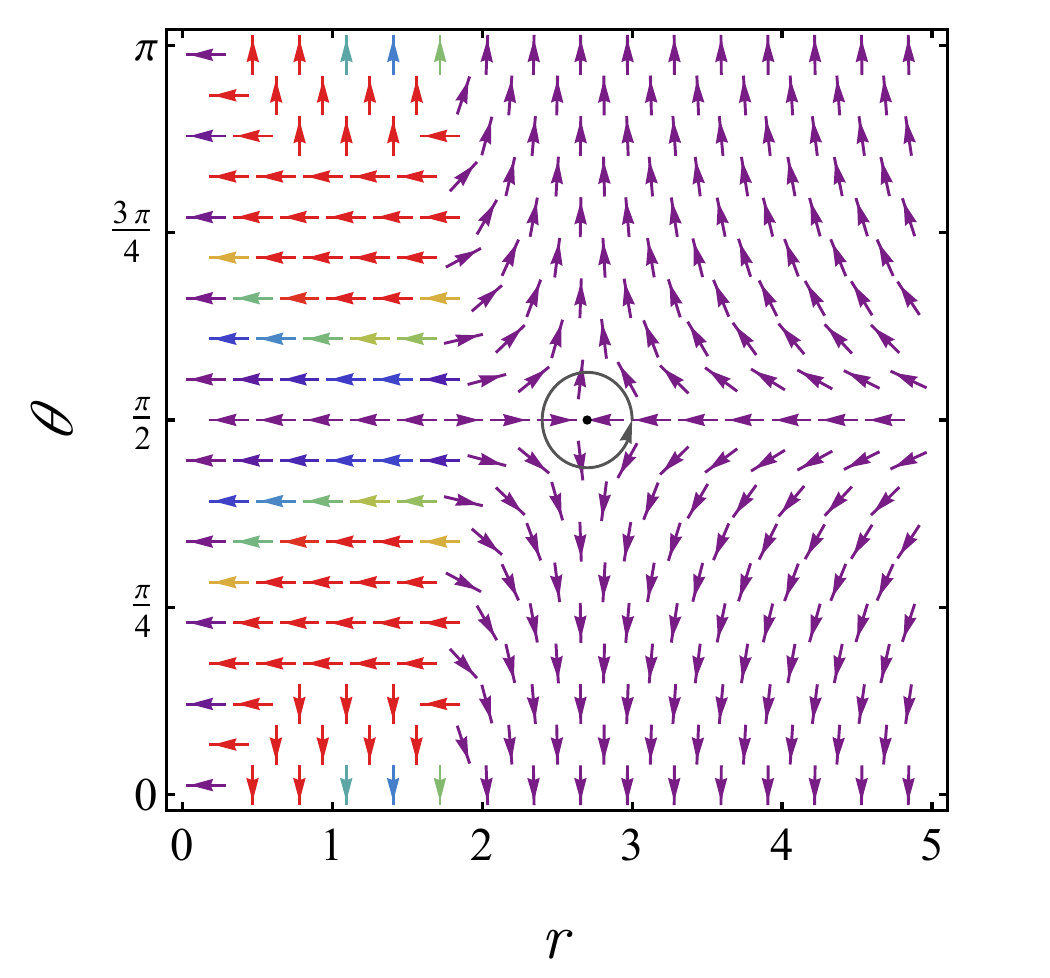}
    \caption{In the $(r, \theta)$ domain, the normalized vector field $n$ is depicted considering the non--commutative framework applied to a Kalb--Ramond black hole. For the parameter set $M = 1$, $\ell = 0.1$, and $\Theta = 0.1$, a well–defined critical point appears at $(r, \theta) = (2.70014, 1.57)$, marking the position of the photon sphere. }
    \label{fig:KR_VectorRph}
\end{figure}

Each zero of the vector field is interpreted as a topological defect corresponding to a photon sphere. The sign of the topological charge characterizes its nature: $ \mathcal{Q} = -1$ in Unstable photon sphere, maximum of potential. $\mathcal{Q} = +1$ in stable photon sphere, minimum of potential.
In the Kalb-Ramond spacetime under consideration, we identify a single topological defect outside the event horizon with winding number $\omega = -1$, indicating an unstable photon sphere. As shown in Fig.~\ref{fig:KR_VectorRph} a single topological defect outside the event horizon is revealed. The winding number for this point encircled by a grey circle is $ \omega = -1$, indicating an unstable photon sphere.

%%%%%%%%%%%%%%%%%%%%%%%%%%%%
%%%%%%%%%%%%%%%%%%%
\subsection{EHT-based constraints on shadow radii}

\begin{figure}
    \centering
     \includegraphics[scale=0.6]{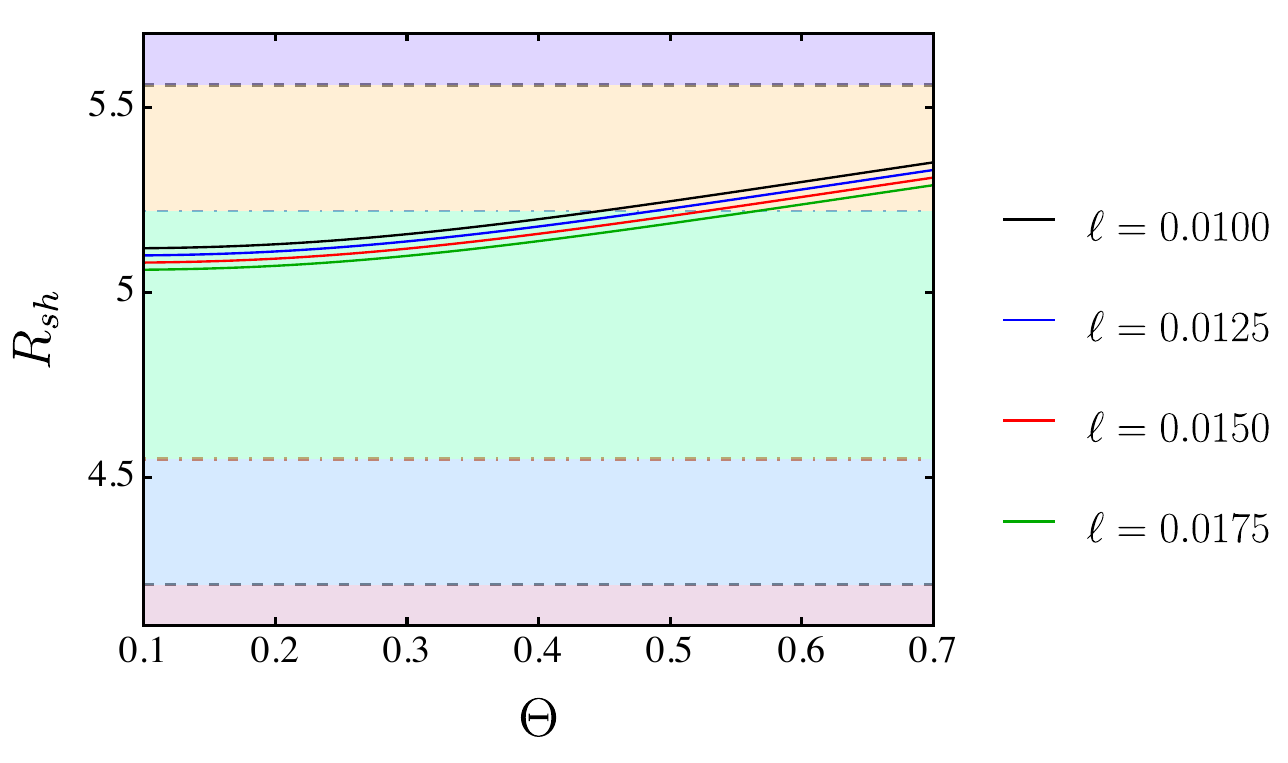}
    \caption{The dependence of the shadow radius on the non--commutative parameter $\Theta$ is examined under the observational bounds associated with $SgrA^{*}$, as reported in \cite{vagnozzi2022horizon,akiyama2022firstSgrA}.}
    \label{adaconstraints}
\end{figure}

\begin{figure}
    \centering
     \includegraphics[scale=0.6]{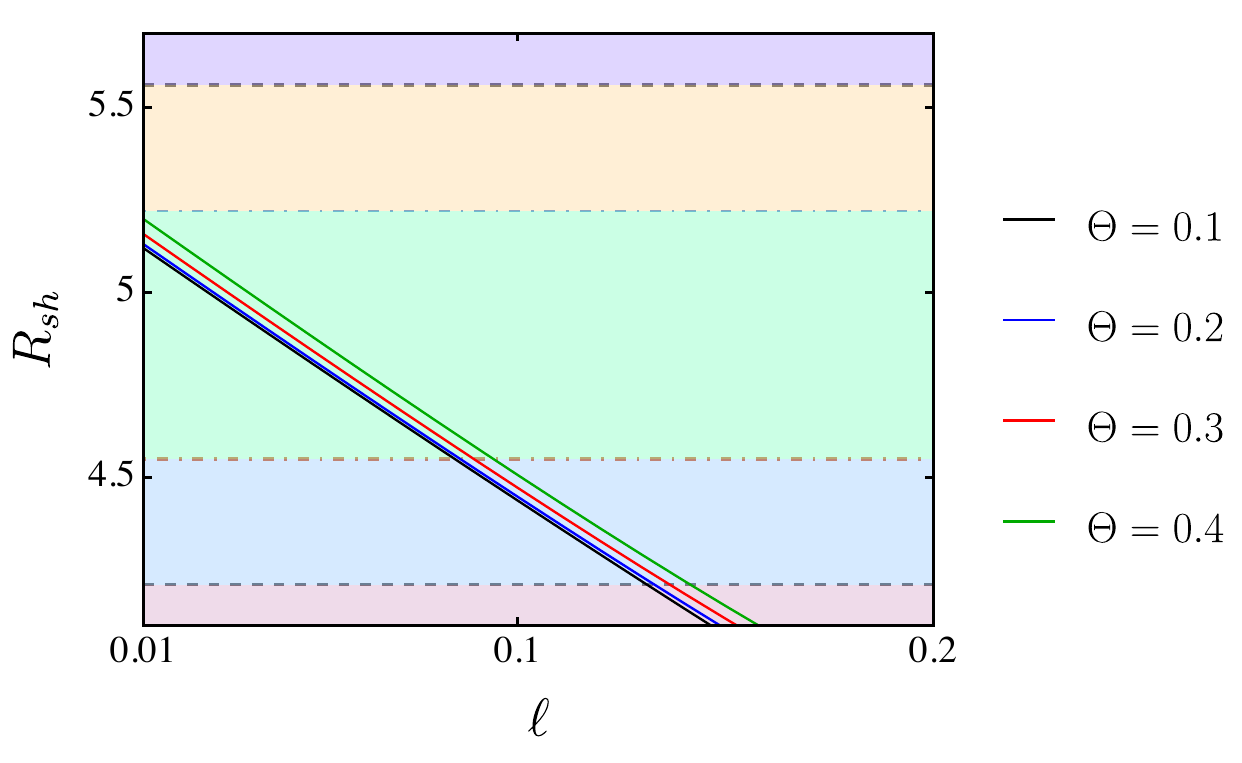}
    \caption{Using the observational data from $SgrA^{*}$ as a reference \cite{vagnozzi2022horizon,akiyama2022firstSgrA}, the dependence of the shadow radius on the Lorentz--violating parameter $\ell$ is explored for a range of fixed values of the non--commutative parameter $\Theta$.}
    \label{adaconstraints2}
\end{figure}

By incorporating horizon--scale observations of $Sgr A^*$ obtained through the Event Horizon Telescope, one can derive constraints on the shadow radius using the independent determinations of the mass--to--distance ratio provided by the VLTI and Keck measurements. When a $2\sigma$ confidence level is considered, as discussed in \cite{heidari2023gravitational}, this approach results in two separate estimates for the shadow size, as outlined in the referenced works \cite{vagnozzi2022horizon,akiyama2022firstSgrA}
\ie
\label{adaconst1asas}
4.55 < \frac{R_{sh}}{M} < 5.22,
\fe
and
\ie
\label{adaconst1asas2}
4.21 < \frac{R_{sh}}{M} < 5.56.
\fe
Theoretical estimates are compared against observational results from the Event Horizon Telescope to infer permissible values for the parameters $\Theta$ and $\ell$. Their influence on the shadow radius—normalized by the black hole mass $M$—is shown in Figs. \ref{adaconstraints} and \ref{adaconstraints2}. The allowed regions in parameter space are identified by locating the intersection between the theoretical curves and the observational uncertainty ranges. Tables \ref{adatab:constr} and \ref{adatab:constr2} compile all corresponding bounds.

The bounds on $\Theta$ vary depending on the choice of $\ell$. When $\ell$ is fixed at $0.0100$, the upper limit for $\Theta$ reaches approximately $0.45$, while for $\ell = 0.0125$, the corresponding upper bound rises to about $0.49$. Increasing $\ell$ to $0.0150$ yields $\Theta \lesssim 0.53$, and with $\ell = 0.0175$, the threshold extends to nearly $0.57$. Conversely, setting specific values for $\Theta$ allows us to delimit the admissible range for $\ell$. For instance, if $\Theta = 0.1$, then $\ell$ is constrained within $0.0793 \lesssim \ell \lesssim 0.129$; for $\Theta = 0.2$, the range shifts to $0.0901 \lesssim \ell \lesssim 0.130$; at $\Theta = 0.3$, the interval becomes $0.0903 \lesssim \ell \lesssim 0.139$; and for $\Theta = 0.4$, the viable window narrows slightly to $0.0905 \lesssim \ell \lesssim 0.140$.

\begin{table}[h!]
\centering
\caption{Limits on the non--commutative parameter $\Theta$ are established by confronting theoretical predictions with the Event Horizon Telescope observations of $SgrA^{*}$ \cite{vagnozzi2022horizon,akiyama2022firstSgrA}.}
\label{adatab:constr2}
\begin{tabular}{lc}
\hline\hline
\textbf{Parameter} & Bounds  \\
\hline
\quad  $\ell =0.0100$  \, \, &  \makecell{$  \Theta \lesssim 0.45$} \\
\quad  $\ell = 0.0125$  \, \,   & \makecell{$\Theta \lesssim 0.49$}  \\
\quad  $\ell = 0.0150$ \, \,  & \makecell{$\Theta \lesssim 0.53$}  \\
\quad $\ell = 0.0175 $ \, \, & \makecell{$\Theta \lesssim  0.57  $}  \\
\hline\hline
\end{tabular}
\end{table}

\begin{table}[h!]
\centering
\caption{Constraints on the Lorentz--violating parameter $\ell$ are derived by comparing theoretical shadow predictions with the observational data from the Event Horizon Telescope for $SgrA^{*}$ \cite{vagnozzi2022horizon,akiyama2022firstSgrA}.}
\label{adatab:constr}
\begin{tabular}{lc}
\hline\hline
\textbf{Parameter} & Bounds  \\
\hline
\quad  $\Theta = 0.1$  & \makecell{ $0.0793 \lesssim \ell \lesssim 0.129 $} \\
\quad  $\Theta = 0.2$     & \makecell{$0.0901 \lesssim \ell \lesssim 0.130 $}  \\
\quad  $\Theta = 0.3$   & \makecell{$0.0903 \lesssim \ell \lesssim 0.139$}  \\
\quad $\Theta = 0.4$  & \makecell{$0.0905 \lesssim \ell \lesssim 0.140 $}  \\
\hline\hline
\end{tabular}
\end{table}

%%%%%%%%%%%%%%%%%%%%%%%%%%%%%%%%%%%%%%%%%%%%%%%%%%%%%%%%%%%%%%%%%%%%%%%%%%%%%%%%%%%%%%%%%%%%%%%%%%%%%%%%%%%%%%%%%%%%%%%%%%%%%%%%%%%%%%%%%%%%%%%%%%%%%%%%%%%%%%%%%%%%%%%%%%%%%%%%%%%%%%%%%%%%%%%%%%%%%%%%%%%%%%%%%%%%%%%%%%%%%%%%%%%%%%%%%%%%%%%%%%%%%%%%%%%%%%%%%%%%%%%%%%%%%%%%%%%%%%%%%%%%%%%%%%%%%%%%%%%%%%%%%%%%%%%%%%%%%%%%%%%%%%%%%%%%%%%%%%%%%%%%%%%%%%%%%%%%%%%%%%%%%%%%%%%%%%%%%%%%%%%%%%%%%%%%%%%%%%%%%%%%%%%%%%%%%%%%%%%%%%%%%%%%%%%%%%%%%%%%%%%%%%%%%%%%%%%%%%%%%%%%%%%%%%%%%%%%%%%%%%%%%%%%%%%%%%%%%%%%%%%%%%%%%%%%%%%%%%%%%%%%%%%%%%%%%%%%%%%%%%%%%%%%%%%%%%%%%%%%%%%%%%%%%%%%%%%%%%%%

%%%%%%%%%%%%%%%%%%%%%%%%%%%%%
%%%%%%%%%%%%%%%%%%%%%%%%%%%%%%
\section{\label{Sec10}Light deflection in a weak gravitational field}

This part of the analysis focuses on the Gaussian curvature $\mathcal{K}(r,\ell,\Theta)$, an essential feature for providing the dynamical stability of photon orbits. The deflection angle associated with gravitational lensing will be computed via the Gauss--Bonnet theorem, adopting the framework established in Ref. \cite{Gibbons:2008rj}. As we shall be seeing, the results indicate that the photon spheres considered here exhibit instability, as reflected by the negative curvature values, $\mathcal{K}(r,\ell,\Theta) < 0$. Furthermore, as the non--commutative parameter $\Theta$ increases, the resulting deflection angle $\hat{\alpha}(b,\ell,\Theta)$ will become more substantial. In contrast, increasing $\ell$ produces the opposite effect, leading therefore to a reduction in the deflection angle.

%%%%%%%%%%%%%%%%%%%%%%%%%%%%%%%%%%%%%%%%%%%%%%%%%%%%%%%%%%%%%%%%%%%%%%%%%%%%%%%%%%%%%%%%%%%%%%%%%%%%%%%%%%%%%%%%%%%%%%%%%%%%%%%%%%%%%%%%%%%%%%%%%%%%%%%%%%%%%%%%%%%%%%%%%%%%%%%%%%%%%%%%%%%%%%%%%%%%%%%%%%%%%%%%%%%%%%%%%%%%%%%%%%%%%%%%%%%%%%%%%%%%%%%%%%%%%%%%%%%%%%%%%%%%%%%%%%%%%%%%%%%%%%%%%%%%%%%%%%%%%%%%%%%%%%%%%%%%%%%%%%%%%%%%%%%%%%%%%%%%%%%%%%%%%%%%%%%%%%%%%%%%%%%%%%%%%%%%%%%%%%%%%%%%%%%%%%%%%%%%%%%%%%%%%%%%%%%%%%%%%%%%%%%%%%%%%%%%%%%%%%%%%%%%%%%%%%%%%%%%%%%%%%%%%%%%%%%%%%%%%%%%%%%%%%%%%%%%%%%%%%%%%%%%%%%%%%%%%%%%%%%%%%%%%%%%%%%%%%%%%%%%%%%%%%%%%%%%%%%%%%%%%%%%%%%%%%%%%%%%

\subsection{Analyzing photon sphere stability via Gaussian curvature}

The behavior of light rays near black holes is closely linked to the structure of the underlying optical geometry, where features such as conjugate points help determine the nature of photon orbits. The response of these orbits to small perturbations reveals whether the associated photon sphere is stable or not. For unstable configurations, even minimal deviations in a photon's path result in either capture by the black hole or escape toward spatial infinity. In contrast, when the photon sphere is stable, such deviations allow photons to remain confined in bounded trajectories around the black hole \cite{qiao2022curvatures,qiao2022geometric}.

Whether conjugate points exist within the spacetime manifold significantly influences the behavior of photon orbits. Stable photon spheres are characterized by the presence of such points, while their absence signals instability. This distinction is supported by the Cartan--Hadamard theorem, which correlates the sign of the Gaussian curvature $\mathcal{K}(r)$ to the existence of conjugate points. In other words, it offers a geometric criterion for evaluating the stability of critical null trajectories \cite{qiao2024existence}. In line with this perspective, null geodesics—defined by the condition $\mathrm{d}s^2 = 0$—can be written as \cite{araujo2024effects,heidari2024absorption}:
\ie
\mathrm{d}t^2=\gamma_{ij}\mathrm{d}x^i \mathrm{d}x^j = \frac{B(\Theta,\ell)}{A(\Theta,\ell)}\mathrm{d}r^2  +\frac{\Bar{D}(\Theta,\ell)}{A(\Theta,\ell)}\mathrm{d}\varphi^2   ,
\fe
with the indices $i$ and $j$ run from $1$ to $3$, and $\gamma_{ij}$ denotes the components of the optical metric. The term $\Bar{D}(\Theta,\ell)$ stands for the function $D(\Theta,\ell)(r,\theta)$ evaluated specifically at the equatorial plane, i.e., $\theta = \pi/2$. Furthermore, the expression for the Gaussian curvature relevant to the optical geometry is provided in Ref. \cite{qiao2024existence}:
\ie
\mathcal{K}(r,\ell,\Theta) = \frac{R}{2} =  - \frac{A(\Theta,\ell)}{\sqrt{B(\Theta,\ell) \,  \Bar{D}(\Theta,\ell)}}  \frac{\partial}{\partial r} \left[  \frac{A(\Theta,\ell)}{2 \sqrt{B(\Theta,\ell) \, \Bar{D}(\Theta,\ell) }}   \frac{\partial}{\partial r} \left(   \frac{\Bar{D}(\Theta,\ell)}{A(\Theta,\ell)}    \right)    \right].
\fe

In the regime where both $\ell$ and $\Theta$ remain small, an explicit form can be obtained for the two--dimensional Ricci scalar, denoted by $R$. Under these conditions, the expression simplifies to:
\ie
\begin{split}
\label{gaussiancurvature}
\mathcal{K}(r,\ell,\Theta)  &  \, \,  \approx  \,  \frac{3 M^2}{r^4} - \frac{2 M}{r^3}  -\frac{2 \ell M}{r^3}-\frac{260 \Theta ^2 \ell M^4}{r^6 (r-2 M)^2} +\frac{358 \Theta ^2 \ell M^3}{r^5 (r-2 M)^2}-\frac{325 \Theta ^2 \ell M^2}{2 r^4 (r-2 M)^2} \\
&   +\frac{51 \Theta ^2 \ell M}{2 r^3 (r-2 M)^2}-\frac{\Theta ^2 \ell}{2 r^2 (r-2 M)^2}+\frac{624 \Theta ^2 M^4}{8 M r^7-4 r^8}-\frac{848 \Theta ^2 M^3 r}{8 M r^7-4 r^8} \\
& +\frac{374 \Theta ^2 M^2 r^2}{8 M r^7-4 r^8}+\frac{\Theta ^2 r^4}{8 M r^7-4 r^8}-\frac{54 \Theta ^2 M r^3}{8 M r^7-4 r^8}  .
\end{split}
\fe
The initial pair of terms in the Gaussian curvature expression are associated with the standard Schwarzschild geometry. The subsequent term arise from modifications due to the Kalb--Ramond background. The remaining contributions highlights the effects of non--commutative geometry alone or in conjunction with the Kalb--Ramond background.

\begin{figure}
    \centering
    \includegraphics[scale=0.65]{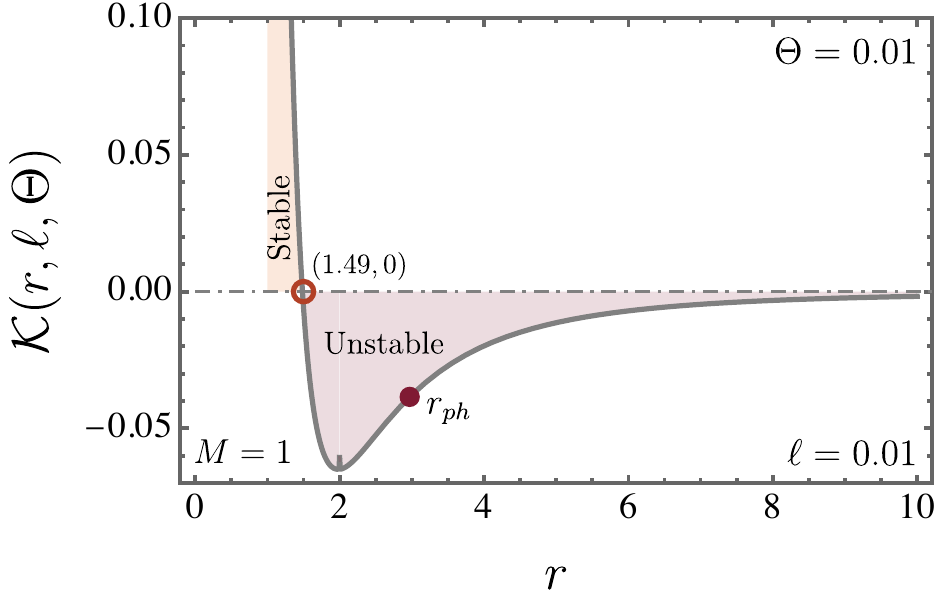}
    \caption{The profile of the Gaussian curvature $\mathcal{K}(r,\ell,\Theta)$ is shown, clearly distinguishing between regions corresponding to stable and unstable photon trajectories. The analysis adopts the parameter values $M = 1$, $\ell = 0.001$, and $\Theta = 0.01$ throughout the plot.}
    \label{figgauss}
\end{figure}

According to Refs. \cite{qiao2022curvatures, qiao2022geometric, qiao2024existence}, the stability of photon orbits can be inferred from the sign of the Gaussian curvature $\mathcal{K}(r,\ell,\Theta)$. When $\mathcal{K}$ is negative, the photon sphere is unstable; conversely, a positive value signals stability. To visualize this behavior, Fig. \ref{figgauss} plots $\mathcal{K}(r,\ell,\Theta)$ as a function of the radial coordinate $r$, delineating regions associated with unstable and stable photon configurations. The plot corresponds to the parameter choices $M = 1.0$, $\ell = 0.01$, and $\Theta = 0.01$. The point at which the curvature changes sign—marking the boundary between stability and instability—is located at $(1.49, 0)$ and indicated by a wine-colored marker (small circle). For this specific setup, the photon sphere radii are approximately $r_{ph} \approx 2.9704$ (wine dot). Furthermore, notice that it is situated in regions where $\mathcal{K}(r,\ell,\Theta) < 0$, confirming that these orbits are unstable therefore.

%%%%%%%%%%%%%%%%%%%%%%%%%%%%%%%%%%%%%%%%%%%%%%%%%%%%%%%%%%%%%%%%%%%%%%%%%%%%%%%%%%%%%%%%%%%%%%%%%%%%%%%%%%%%%%%%%%%%%%%%%%%%%%%%%%%%%%%%%%%%%%%%%%%%%%%%%%%%%%%%%%%%%%%%%%%%%%%%%%%%%%%%%%%%%%%%%%%%%%%%%%%%%%%%%%%%%%%%%%%%%%%%%%%%%%%%%%%%%%%%%%%%%%%%%%%%%%%%%%%%%%%%%%%%%%%%%%%%%%%%%%%%%%%%%%%%%%%%%%%%%%%%%%%%%%%%%%%%%%%%%%%%%%%%%%%%%%%%%%%%%%%%%%%%%%%%%%%%%%%%%%%%%%%%%%%%%%%%%%%%%%%%%%%%%%%%%%%%%%%%%%%%%%%%%%%%%%%%%%%%%%%%%%%%%%%%%%%%%%%%%%%%%%%%%%%%%%%%%%%%%%%%

\subsection{Applying the Gauss–Bonnet theorem for addressing gravitational lensing}

Starting from the expression for the Gaussian curvature given in Eq. (\ref{gaussiancurvature}), we proceed to apply the Gauss--Bonnet theorem to compute the deflection angle under the weak--field approximation \cite{Gibbons:2008rj}. In this manner, it is required the calculation of the so--called optical surface integral restricted to the equatorial plane, which is formulated as follows:
\ie
\mathrm{d}S = \sqrt{\gamma} \, \mathrm{d} r \mathrm{d}\varphi = \sqrt{\frac{B(\Theta,\ell)}{A(\Theta,\ell)}  \frac{D(\Theta,\ell)}{A(\Theta,\ell)} } \, \mathrm{d} r \mathrm{d}\varphi,
\fe
which in turn leads to the deflection angle being obtained by
\ie
\begin{split}
& \hat{\alpha} (b,\ell,\Theta) =  - \int \int_{D} \mathcal{K} \mathrm{d}S = - \int^{\pi}_{0} \int^{\infty}_{\frac{b}{\sin \varphi}} \mathcal{K} \mathrm{d}S \\
 \simeq  & \, \, \frac{4 M}{b} +\frac{ {15} \pi  M^2}{4 b^2}-\frac{2 \ell M}{b} -\frac{ {21} \pi  \ell M^2}{8 b^2} -\frac{ {-5} \Theta ^2 \ell M}{ {3} b^3}-\frac{ {14}\, \Theta ^2 M}{3 b^3} \\
& -\frac{ {587} \pi  \Theta ^2 \ell M^2}{ {128} b^4} {-} \frac{{839}  \pi  \Theta ^2 M^2}{ {256} b^4}+\frac{\pi  \Theta ^2}{16 b^2}+\frac{\pi  \Theta ^2 \ell}{32 b^2}.
\end{split}
\fe

\begin{figure}
    \centering
    \includegraphics[scale=0.55]{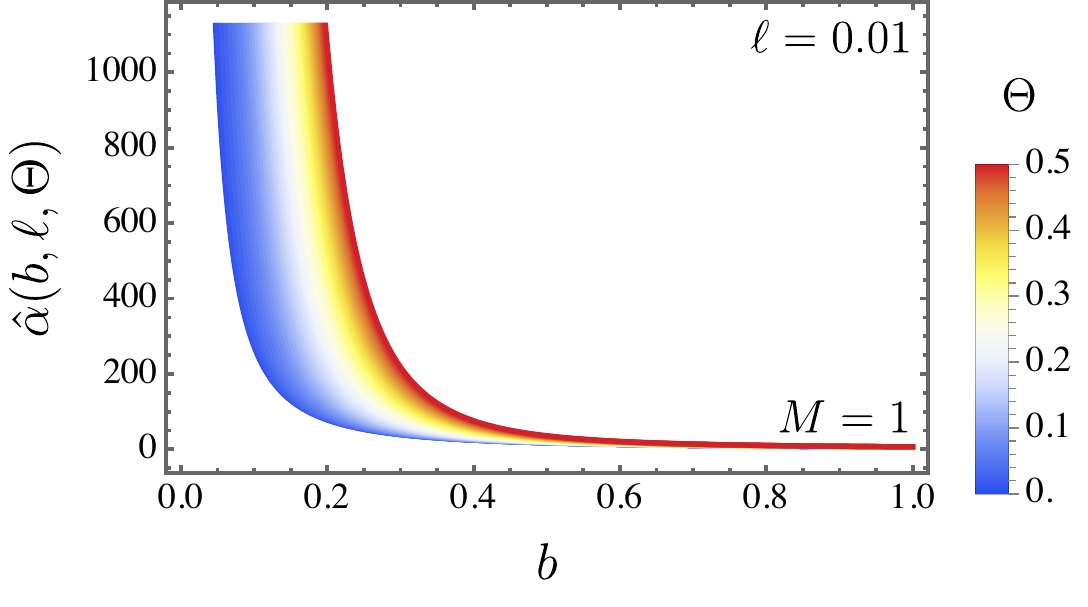}
    \includegraphics[scale=0.55]{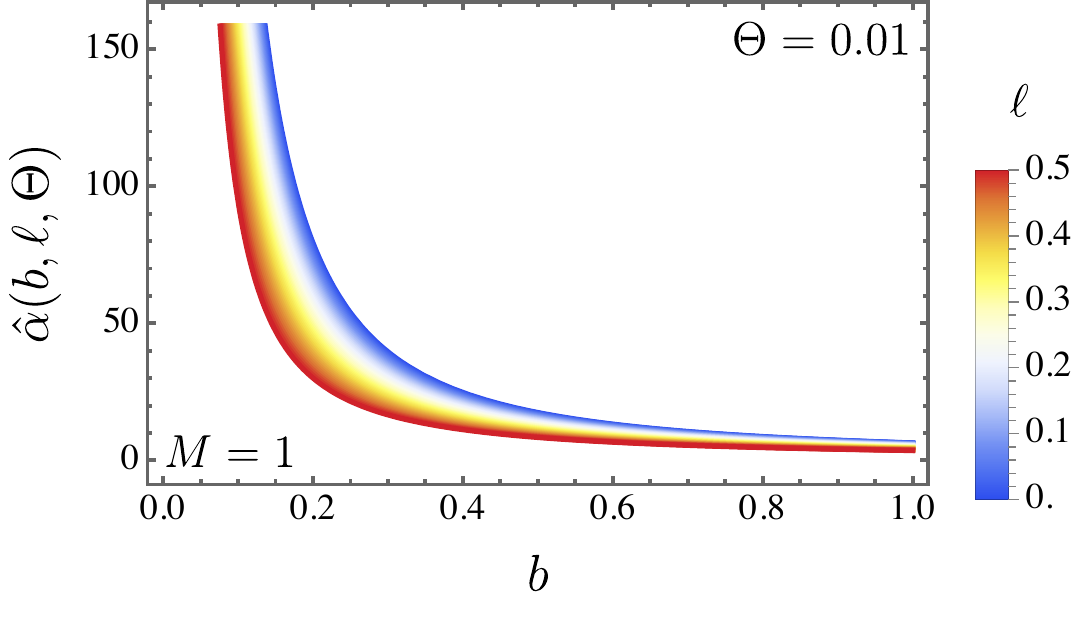}
    \caption{The top panel illustrates how the deflection angle $\hat{\alpha}(b,\ell,\Theta)$ varies with the impact parameter $b$ for different values of $\Theta$, while keeping the Lorentz-violating parameter fixed at $\ell = 0.01$. In contrast, the bottom panel displays the behavior of $\hat{\alpha}(b,\ell,\Theta)$ for varying $\ell$ values, with $\Theta$ held constant at $0.01$.}
    \label{a1n1g1l1d1e1f1c}
\end{figure}

Fig. \ref{a1n1g1l1d1e1f1c} presents how the deflection angle $\hat{\alpha}(b,\ell,\Theta)$ is affected by the deformation parameters. {The first two terms reproduce the standard Schwarzschild contribution, while the third and fourth terms account for the effects introduced by the Kalb–Ramond background. The remaining terms arise from the non--commutative deformation or a combination of non--commutativity and the Kalb--Ramond corrections proposed in this work.}

For a fixed impact parameter, such as $b = 0.2$, increasing $\Theta$ leads to a more pronounced deflection. In contrast, larger values of $\ell$ tend to suppress the deflection angle, resulting, therefore, in a reduced value of $\hat{\alpha}(b,\ell,\Theta)$.

%%%%%%%%%%%%%%%%%%%%%%%%%%%%%%%%%%%%%%%%%%%%%%%%%%%%%%%%%%%%%%%%%%%%%%%%%%%%%%%%%%%%%%%%%%%%%%%%%%%%%%%%%%%%%%%%%%%%%%%%%%%%%%%%%%%%%%%%%%%%%%%%%%%%%%%%%%%%%%%%%%%%%%%%%%%%%%%%%%%%%%%%%%%%%%%%%%%%%%%%%%%%%%%%%%%%%%%%%%%%%%%%%%%%%%%%%%%%%%%%%%%%%%%%%%%%%%%%%%%%%%%%%%%%%%%%%%%%%%%%%%%%%%%%%%%%%%%%%%%%%%%%%%%%%%%%%%%%%%%%%%%%%%%%%%%%%%%%%%%%%%%%%%%%%%%%%%%%%%%%%%%%%%%%%%%%%%%%%%%%%%%%%%%%%%%%%%%%%%%%%%%%%%%%%%%%%%%%%%%%%%%%%%%%%%%%%%%%%%%%%

\section{Constraining Non-Commutativity via EHT Observations}

Recent observations captured by the Event Horizon Telescope (EHT) of the supermassive black hole $Sgr A^*$ \cite{akiyama2022firstSgr,akiyama2022firstSgrA} offer a valuable opportunity to probe deviations from classical general relativity, particularly within the context of non-commutative geometry. In this work, we utilize these observations to assess potential bounds on the non--commutative parameter $\Theta$, considering a static Kalb--Ramond black hole embedded in a non--commutative spacetime.

One of the key observables accessible through EHT measurements is the angular diameter of the shadow, denoted by $\Omega_{\text{sh}}$. This quantity is sensitive to modifications in the underlying gravitational theory, including those arising from quantum gravity corrections \cite{afrin2023tests}. By confronting theoretical predictions for $\Omega_{\text{sh}}$ with empirical values reported by the EHT, one can place meaningful constraints on $\Theta$.

For an observer positioned at a distance $d_O$ from the black hole, the angular size of the shadow is determined by \cite{kumar2020rotating}:
\begin{equation}
\Omega_{\text{sh}} = \frac{2b_c}{d_O},
\end{equation}

Here, $b_c$ represents the critical impact parameter that corresponds to the photon sphere. Expressed in observational units, this becomes

\cite{xu2025optical,heydari2024effect}
\begin{equation}
\Omega_{\text{sh}}  = \frac{6.191165 \times 10^{-8}}{\pi} \, \frac{\gamma}{\frac{d_O} {\textit{Mpc}}} \left(\frac{b_{\text{c}}}{M}\right)\mu\text{as}
\end{equation}
with \(\gamma = M / M_{\odot}\) being the mass in solar units.
In the case of $Sgr A^*$, the EHT collaboration reported a black hole mass of $M = 4 \times 10^6 M_{\odot}$ and an observer distance of $d_O = 8.15$ kpc. The angular diameter of the shadow is estimated to be $\Omega_{\text{sh}} = 48.7 \pm 7\mu\text{as}$ \cite{akiyama2022firstSgr,akiyama2022firstSgrA}. 
We use these observational data to evaluate how the non--commutativity impacts the predicted angular shadow size, and whether consistency with the EHT data imposes a meaningful constraints on Lorenz--violating parameter $\ell$, or non--commutativity $\Theta$.

\begin{figure}[ht]
    \centering
    \includegraphics[width=90mm]{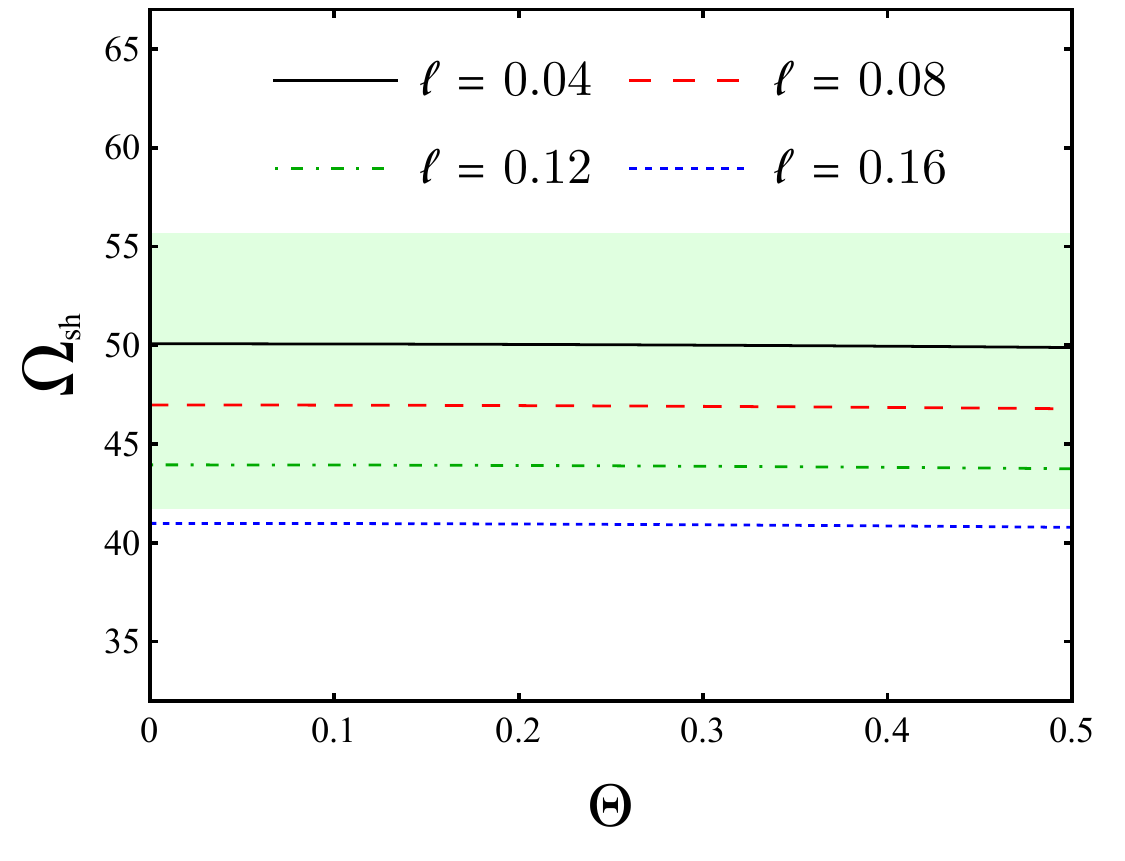}
    \caption{The angular diameter of the shadow $\Omega_{\text{sh}}$ as a function of the Kalb--Ramond parameter $\ell$ for fixed values of the non--commutative parameter $\Theta$. The green region represents interval of $Sgr A^*$'s angular shadow diameter.}
    \label{fig:SgrAL}
\end{figure}

\begin{figure}[ht]
    \centering
    \includegraphics[width=90mm]{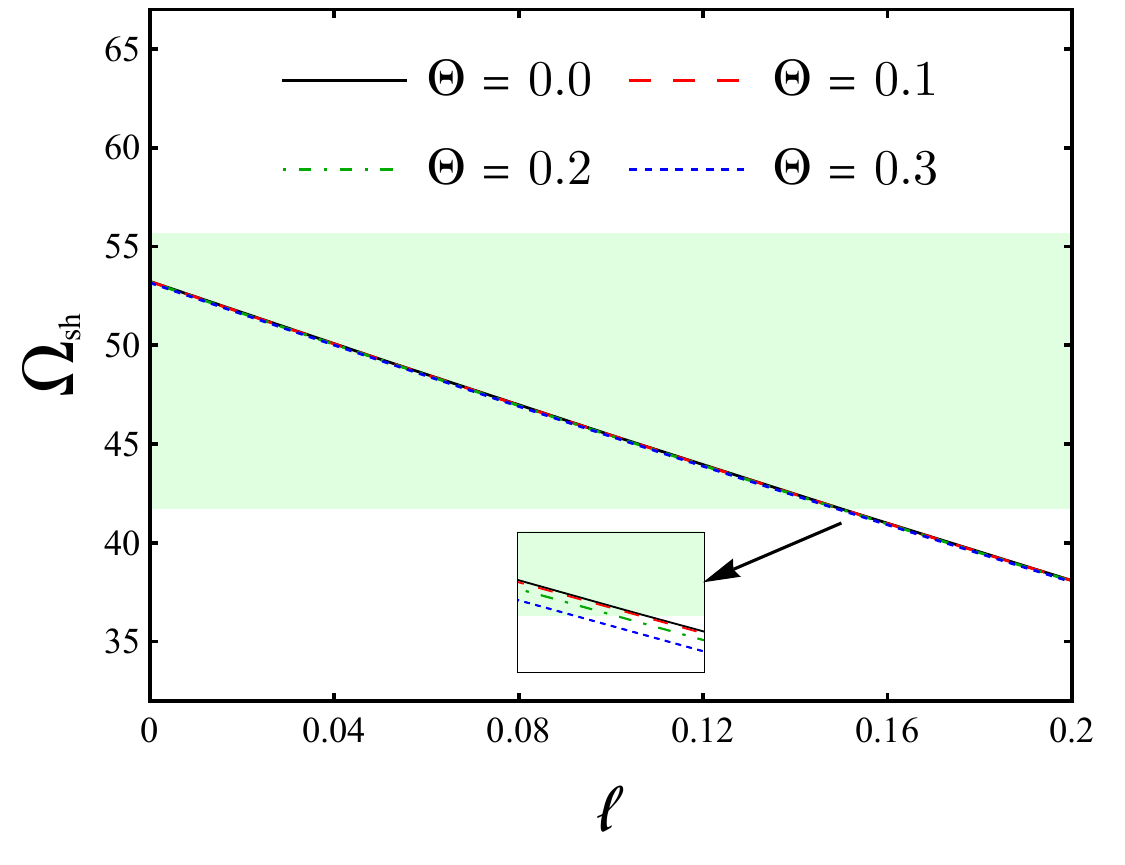}
    \caption{Dependence of $\Omega_{\text{sh}}$ on the non--commutative parameter $\Theta$ for selected values of $\ell$. The interval of $Sgr A^*$'s angular shadow diameter by EHT data is demonstrated by green region. }
    \label{fig:SgrATheta}
\end{figure}

Fig.~\ref{fig:SgrAL} and \ref{fig:SgrATheta} display the dependence of \(\Omega_{\text{sh}}\) on the Kalb--Ramond parameter \(\ell\) and the noncommutative parameter \(\Theta\). The green bands represent the EHT observational range for the angular shadow diameter. This comparison allows us to constrain the admissible ranges of \(\ell\) and \(\Theta\), based on their consistency with the observed angular shadow size.

%%%%%%%%%%%%%%%%%%%%%%%%%%%%%%%%%%%%

Fig.~\ref{fig:SgrAL} illustrates the behavior of the angular shadow diameter $\Omega_{\text{sh}}$ with respect to the non--commutative parameter $\Theta$ for various fixed values of the Kalb--Ramond parameter $\ell$. The results indicate that $\Theta$ induces only subtle changes in $\Omega_{\text{sh}}$, with the curves remaining nearly flat and within the observational bounds provided by the EHT collaboration ($48.7 \pm 7\,\mu\text{as}$) based on the $\ell$ values.

Instead, the impact of the Kalb--Ramond parameter $\ell$ is more significant. As $\ell$ increases, the entire $\Omega_{\text{sh}}$ curve shifts downward. Notably, for larger values of $\ell$, the predicted angular shadow diameter exceeds the lower observational limit across all values of $\Theta$, rendering these parameter combinations incompatible with EHT measurements. This behavior implies the existence of a strict upper bound on $\ell$ beyond which the model predictions violate empirical constraints.

To quantify the allowed parameter space, we examine $\Omega_{\text{sh}}$ explicitly as a function of $\ell$ in Fig.~\ref{fig:SgrATheta}, to check the constraints on $\ell$ for different values of $\Theta$. According to this figure, for all choices of $\Theta$, the behavior of $\Omega_{\text{sh}}$ with respect to $\ell$ exhibits a similar trend and $\Omega_{\text{sh}}$ monotonically decreases as $\ell$ increases.  However, beyond a critical threshold of $\ell$, the predicted angular shadow diameter $\Omega_{\text{sh}}$ falls below the observational lower bound ($41.7 \mu\text{as}$), rendering those parameter values observationally excluded. Notably, the maximum allowed value of $\ell$---i.e., the upper bound ensuring $\Omega_{\text{sh}}$ remains consistent with EHT constraints---depends on $\Theta$. As $\Theta$ increases, this upper bound on $\ell$ sligtly decreases. This correlation between $\ell$ and $\Theta$ emphasize a non--trivial result as the larger non--commutative effects $\Theta$ require a smaller Lorentz--violating parameter $\ell$ to maintain agreement with the observed angular shadow size of $Sgr A^*$.

%%%%%%%%%%%%%%%%%%%%%%%%%%%%%%%%%%%%%%%%%%%%%%%%%%%%%%%%%%%%%%%%%%%%%%%%%%%%%%%%%%%%%%%%%%%%%%%%%%%%%%%%%%%%%%%%%%%%%%%%%%%%%%%%%%%%%%%%%%%%%%%%%%%%%%%%%%%%%%%%%%%%%%%%%%%%%%%%%%%%%%%%%%%%%%%%%%%%%%%%%%%%%%%%%%%%%%%%%%%%%%%%%%%%%%%%%%%%%%%%%%%%%%%%%%%%%%%%%%%%%%%%%%%%%%%%%%%%%%%%%%%%%%%%%%%%%%%%%%%%%%%%%%%%%%%%%%%%%%%%%%%%%%%%%%%%%%%%%%%%%%%%%%%%%%%%%%%%%%%%%%%%%%%%%%%%%%%%%%%%%%%%%%%%%%%%%%%%%%%%%%%%%%%%%%%%%%%%%%%%%%%%%%%%%%%%%%%%%%%%%

\section{Strong field limit of light deflection \label{stronggggsss}}

{%
This section investigates gravitational lensing in the strong deflection limit (SDL) for the Kalb--Ramond black hole described by Eq.~(\ref{metrictensorss}).  Because the non--commutative correction to the metric component $D$ in Eq.~(\ref{metrictensorss}) explicitly depends on the polar angle $\theta$, the spacetime is axisymmetric rather than spherically symmetric.  The original SDL formalism of Tsukamoto~\cite{035}, designed for the most general spherically symmetric geometry, therefore cannot be applied directly.  We instead employ the recent coordinate--independent generalization introduced by Igata~\cite{Igata:2025hpy,Igata:2025taz}, which encompasses arbitrary stationary, axisymmetric spacetimes.  Since the Kalb--Ramond metric (\ref{metrictensorss}) is static, all quantities associated with co--rotating and counter--rotating photon orbits coincide, and the formalism reduces to a single branch.
}

%%%%%%%%%%%%%%%%%%%%%%%%%%%%%%%%%%%%%%%%%%%%%%%%%%%%%%%%%%%%%%%%%%%%%%%%%%%%%%%%%%%%%%%%%%%%%%%%%%%%%%%%%%%%%%%%%%%%%%%%%%%%%%%%%%%%%%%%%%%%%%%%%%%%%%%%%%%%%%%%%%%%%%%%%%%%%%%%%%%%%%%%%%%%%%%%%%%%%%%%%%%%%%%%%%%%%%%%%%%%%%%%%%%%%%%%%%%%%%%%%%%%%%%%%%%%%%%%%%%%%%%%%%%%%%%%%%%%%%%%%%%%%%%%%%%%%%%%%%%%%%%%%%%%%%%%%%%%%%%%%%%%%%%%%%%%%%%%%%%%%%%%%%%%%%%%%%%%%%%%%%%%%%%%%%%%%%%%%%%%%%%%%%%%%%%%%%%%%%%%%%%%%%%%%%%%%%%%%%%%%%%%%%%%%%%%%%%%%%%%%%%%%%%%%%%%%%%%%%%%%%%%%%%%%%%%%%%%%%%%%%%%%%%%%%%%%%%%%%%%%%%%%%%%%%%%%%%%%%%%%%%%%%%%%%%%%%%%%%%%%%%%%%%%%%%%%%%%%%%%%%%%%%%%%%%%%%%%%%%%%%%%%%%%%%%%%%%%%%%%%%%%%%%%%%%%%%%%%%%%%%%%

\subsection{Theoretical framework}

The analysis is conducted within a stationary, axisymmetric spacetime characterized by two commuting Killing vectors: $\partial_t$, associated with invariance under time translations, and $\partial_\varphi$, corresponding to axial symmetry. Under these symmetry conditions, the spacetime metric $g_{ab}$ can be expressed in the canonical Weyl--Papapetrou representation~\cite{Stephani:2003tm}
\begin{align}
\label{eq:metric}
\mathrm{d}s^2 =
-\sigma e^{2\psi}\left(\mathrm{d}t+\sigma A\mathrm{d}\varphi\right)^2
+ e^{-2\psi}\left[
e^{2\gamma}\left(\mathrm{d}\rho^2+\mathrm{d}\zeta^2\right)
+\sigma\rho^2\mathrm{d}\varphi^2\right].
\end{align}

{The functions $\psi$, $\gamma$, and $A$ depend solely on the Weyl coordinates $(\rho,\zeta)$; the constant $\sigma = +1$ ($-1$) outside (inside) the static limit surface.  The circumferential radius is $R = e^{-\psi}\sqrt{\sigma(\rho^2-e^{4\psi}A^{2})}$, and a reflection symmetry about the equatorial plane $\zeta = 0$ is assumed.  A zero angular momentum observer (ZAMO)~\cite{Bardeen:1972fi} has lapse $\alpha = \rho/R$ and angular velocity $\Omega = e^{2\psi}A/R^2$ relative to infinity.
}

{The detailed construction of the ZAMO vierbein, Newman--Penrose null tetrad, and Weyl scalar definitions appearing in the original version has been condensed; see Ref.~\cite{Igata:2025hpy} for the complete formalism.}

{Projecting the Weyl tensor $C_{abcd}$ onto the ZAMO tetrad yields the electric and magnetic parts $E_{ij} = C_{i0j0}$ and $B_{ij} = \tfrac{1}{2}C_{i0}{}^{kl}\epsilon_{klj0}$, which encode, respectively, the tidal and frame--dragging contributions to the gravitational field.  The Einstein tensor $G_{ab} = R_{ab}-\tfrac{1}{2}\mathcal{R}\,g_{ab}$, projected onto the same frame, captures the matter--energy content.}

{In the general stationary case, a photon orbiting at circumferential radius $R_{\mathrm{m}}^{\pm}$ has two branches ($+$ for prograde, $-$ for retrograde). The second derivative of the effective potential on each branch, which controls the instability of the circular orbit, is expressed in Igata's decomposition as~\cite{Igata:2025hpy}:
\begin{align}
\label{eq:V''EG}
V''_{\mathrm{m},\pm}
=\left(\frac{R_{\mathrm{m}}^{\pm}}{1\pm v_{\mathrm{m}}^{\pm}}\right)^2\left[
G_{00}^{\mathrm{m},\pm}+G_{33}^{\mathrm{m},\pm}-2\left(E_{22}^{\mathrm{m},\pm}-E_{11}^{\mathrm{m}, \pm}\right)\pm \left(4B_{12}^{\mathrm{m},\pm}+ 2G_{03}^{\mathrm{m},\pm}\right)\right],
\end{align}
with 
\begin{align}
v_{\mathrm{m}}^{\pm}=\frac{R_{\mathrm{m}}^{\pm}\,\Omega_{\mathrm{m}}^{\pm}}{\alpha_{\mathrm{m}}^{\pm}}.
\end{align}
Equivalently, using the Newman--Penrose scalars~\cite{Igata:2025hpy}:
\begin{align}
V''_{\mathrm{m},+} = 4\left(\dfrac{R_\mathrm{m}^{+}}{1+v_\mathrm{m}^{+}}\right)^2 \left(\Psi_0^{\mathrm{m},+}+\Phi_{00}^{\mathrm{m},+}\right)
\end{align}
and
\begin{align}
V''_{\mathrm{m},-} = 4\left(\dfrac{R_\mathrm{m}^{+}}{1-v_\mathrm{m}^{+}}\right)^2 \left(\Psi_4^{\mathrm{m},-}+\Phi_{22}^{\mathrm{m},-}\right).
\end{align}
For the present static metric, $A = 0$ in the Weyl--Papapetrou form, hence $\Omega = 0$ and $v_{\mathrm{m}}^{\pm} = 0$, the magnetic part $B_{ij}$ and the mixed Einstein component $G_{03}$ vanish identically, so that the $\pm$ branches coincide. The instability of the photon sphere is then governed entirely by the Weyl--electric part $E_{ij}$ and the diagonal Einstein components $G_{00}$, $G_{33}$.}

We now restrict to the equatorial plane $\theta = \pi/2$, where by reflection symmetry all equatorial photon orbits remain confined. On this plane, the Kalb--Ramond metric~\mbox{(\ref{metrictensorss})} takes the effective diagonal form
\begin{align}
\label{eq:eqmetric}
\mathrm{d}s^2\big|_{\theta=\pi/2} = -A(r)\,\mathrm{d}t^2 + B(r)\,\mathrm{d}r^2 + D(r)\,\mathrm{d}\varphi^2,
\end{align}
where $A(r)$, $B(r)$, and $D(r)$ are the metric components given in Sec.~\mbox{\ref{Sec2}}, evaluated at $\theta=\pi/2$. Note that $D\big|_{\theta=\pi/2} \neq r^2$ due to the $\Theta^2$-correction in Eq.~\mbox{(\ref{metrictensorss})}, reflecting the departure from spherical symmetry. We define the circumferential radius and the auxiliary ratio
\begin{align}
\label{eq:RfDef}
R(r) \equiv \sqrt{D(r)}, \qquad f(r) \equiv \frac{D(r)}{A(r)}.
\end{align}
The conserved energy $E = A\dot{t}$ and angular momentum $L = D\dot{\varphi}$ yield the impact parameter $b = L/E$. Substituting into the null condition $g_{\mu\nu}\dot{x}^{\mu}\dot{x}^{\nu}=0$ and dividing by $\dot{\varphi}^2$, one obtains the orbital equation
\begin{align}
\label{eq:orbit_eq}
\left(\frac{\mathrm{d}r}{\mathrm{d}\varphi}\right)^{2} + V(r) = 0,
\end{align}
with the effective potential
\begin{align}
\label{eq:Veff}
V(r) = \frac{D}{B}\left(1-\frac{f(r)}{b^2}\right).
\end{align}
A circular photon orbit at $r = r_m$ requires $V(r_m) = 0$ and $V'(r_m)=0$ simultaneously. Since $D_m/B_m > 0$, the first condition fixes the critical impact parameter
\begin{align}
\label{eq:bc}
b_c^{\,2} = f_m \equiv \frac{D_m}{A_m}\,.
\end{align}
Differentiating $V$ and using $f/b^2 = 1$ at $r_m$, the second condition reduces to
\begin{align}
\label{eq:fprime}
f'(r_m)=0\implies\frac{A'_m}{A_m} = \frac{D'_m}{D_m}\,.
\end{align}
The orbit is unstable when $V''_m < 0$. Evaluating the second derivative at $r_m$, where both $f/b^2 = 1$ and $f'_m = 0$, yields
\begin{align}
\label{eq:Vpp}
V''_m &= -\frac{D_m}{B_m}\left(\frac{D''_m}{D_m}-\frac{A''_m}{A_m}\right)\notag\\
&=-\dfrac{2\left[2(\ell-1)M+r\right]^2}{\frac{\Theta ^2 (\ell-1) M [3 (\ell-1) M+2 r]}{r^2}+2 (r-\ell r) [2 (\ell-1) M+r]}\notag\\
&\quad\times\Biggl\{\dfrac{\Theta ^2 \left[-2 (\ell-1) M^2+4 (\ell-1) M r+r^2\right]}{4 r [2 (\ell-1) M+r]}+r^2\Biggr\}\notag\\
&\quad\times\Biggl\{
\dfrac{-8 (\ell-1) M r^3-4 \Theta ^2 M [55 (\ell-1) M+12 r]}{2 r^5 [2 (\ell-1) M+r]+\Theta ^2 M r^2 [11 (\ell-1) M+4 r]}\notag\\
&\quad+\frac{\frac{\Theta ^2 (\ell-1) M \left[-4 (\ell-1)^2 M^3-6 (\ell-1) M^2 r-3 M r^2+r^3\right]}{r^3 [2 (\ell-1) M+r]^3}+2}{\frac{\Theta ^2 \left[-2 (\ell-1) M^2+4 (\ell-1) M r+r^2\right]}{4 r [2 (\ell-1) M+r]}+r^2}.
\Biggr\}
\end{align}
The instability criterion is therefore $D''_m/D_m - A''_m/A_m > 0$. When the Kalb--Ramond metric is cast in Weyl--Papapetrou form, this $V''_m$ can be decomposed according to Eq.~\eqref{eq:V''EG}: for the present static geometry the Weyl--electric and Einstein tensor components are the sole contributors.

%%%%%%%%%%%%%%%%%%%%%%%%%%%%%%%%%%%%%%%%%%%%%%%%%%%%%%%%%%%%%%%%%%%%%%%%%%%%%%%%

%%%%%%%%%%%%%%%%%%%%%%%%%%%%%%%%%%%%%%%%%%%%%%%%%%%%%%%%%%%%%%%%%%%%%%%%%%%%%%%%%%%%%%%%%%%%%%%%%%%%%%%%%%%%%%%%%%%%%%%%%%%%%%%%%%%%%%%%%%%%%%%%%%%%%%%%%%%%%%%%%%%%%%%%%%%%%%%%%%%%%%%%%%%%%%%%%%%%%%%%%%%%%%%%%%%%%%%%%%%%%%%%%%%%%%%%%%%%%%%%%%%%%%%%%%%%%%%%%%%%%%%%%%%%%%%%%%%%%%%%%%%%%%%%%%%%%%%%%%%%%%%%%%%%%%%%%%%%%%%%%%%%%%%%%%%%%%%%%%%%%%%%%%%%%%%%%%%%%%%%%%%%%%%%%%%%%%%%%%%%%%%%%%%%%%%%%%%%%%%%%%%%%%%%%%%%%%%%%%%%%%%%%%%%%%%%%%%%%%%%%%%%%%%%%%%%%%%%%%%%%%%%%%%%%%%%%%%%%%%%%%%%%%%%%%%%%%%%%%%%%%%%%%%%%%%%%%%%%%%%
\subsection{Strong-field deflection angle}

Consider a photon incident from infinity with impact parameter $b$. It attains its distance of closest approach at $r = r_0$, determined by the condition $V(r_0) = 0$ (equivalently, $b^2 = f_0 \equiv D_0/A_0$), before being scattered back to infinity. The total accumulated azimuthal angle and the deflection angle are given by
\begin{align}
\label{eq:Idef}
I(r_0) &= 2\int_{r_0}^{\infty}\frac{|\mathrm{d}r|}{\sqrt{-V(r)}}
\end{align}
and
\begin{align}
\alpha(r_0) = I(r_0)-\pi.
\end{align}
Substituting the relation \eqref{eq:Veff}, this becomes
\begin{align}
\label{eq:Isqrt}
I(r_0) &= 2\sqrt{f_0}\int_{r_0}^{\infty}\frac{\sqrt{B}|\mathrm{d}r|}{\sqrt{D(f-f_0)}}.
\end{align}

To extract the logarithmic divergence arising in the strong-deflection limit ($r_0 \to r_m$), we follow Igata~\mbox{\cite{Igata:2025hpy,Igata:2025taz}} and introduce the coordinate-independent variable
\begin{align}
\label{eq:zdef}
z &\equiv 1 - \frac{R_0}{R(r)} = 1 - \sqrt{\frac{D_0}{D(r)}}\,,
\end{align}
such that $z = 0$ at $r = r_0$ and $z \to 1$ as $r\to\infty$.  Using the differential relation $|{\rm d}r| = R^2/(R_0\,|R'|)\,{\rm d}z$ along with $D = R^2 = R_0^2/(1-z)^2$, we transform the integral \mbox{(\ref{eq:Isqrt})} into
\begin{align}
\label{eq:IzP}
I(r_0) &= 2\int_0^1\frac{\mathrm{d}z}{\sqrt{P(z)}}\,,
\end{align}
where
\begin{align}
\label{eq:Pdef}
P(z) &\equiv \frac{(R')^2}{B\,f_0}\,(1-z)^2\,(f - f_0)\,.
\end{align}
The condition $f(r_0) = f_0$ implies $P(0) = 0$, which underlies the divergence. Expanding $P(z) = c_1\,z + c_2\,z^2 + \mathcal{O}(z^3)$ about $z = 0$ yields, after straightforward algebra~\mbox{\cite{Igata:2025hpy}},
\begin{align}
\label{eq:c1}
c_1 &= -\frac{R'_0V'_0}{R_0},
\end{align}
and
\begin{align}
\label{eq:c2}
c_2 = -\frac{V''_0}{2} + \left(\frac{3R'_0}{R_0}-\frac{R''_0}{2R'_0}\right)V'_0.
\end{align}
The dominant divergent contribution is captured by
\begin{align}
\label{eq:IDdef}
I_{\mathrm{D}}(r_0) &= 2\int_0^1\frac{\mathrm{d}z}{\sqrt{c_1\,z + c_2\,z^2}}\,,
\end{align}
which admits the closed-form evaluation~\mbox{\cite{Igata:2025hpy}}
\begin{align}
\label{eq:IDclosed}
I_{\mathrm{D}}(r_0) &= \frac{4}{\sqrt{c_2}}\,\ln\frac{\sqrt{c_1+c_2}+\sqrt{c_2}}{\sqrt{c_1}}\,.
\end{align}
In the limit where $r_0$ approaches the photon sphere radius $r_m$, the condition $V'_0 \to 0$ implies that $c_1 \to 0$, whereas $c_2$ converges to $c_2^m = -V''_m/2 > 0$. Concurrently, the impact parameter behaves as
\begin{align}
\frac{b}{b_c}-1 &\simeq -\frac{B_m\,V''_m}{4D_m}\,(r_0-r_m)^2.
\end{align}
Substituting these leading-order expansions into Eq.~\mbox{\eqref{eq:IDclosed}} and introducing the regular integral $I_R(r_m) \equiv I(r_m) - I_D(r_m)$, we obtain the master formula for the deflection angle in the strong-deflection limit:
\begin{align}
\label{eq:alphaSDL}
\alpha(b) &\approx -\bar{a}\,\ln\left(\frac{b}{b_c}-1\right) + \bar{b},
\end{align}
with the SDL coefficients
\begin{align}
\label{eq:abar}
\bar{a} &= \sqrt{-\frac{2}{V''_m}},
\end{align}
and
\begin{align}
\label{eq:bbar}
\bar{b} &= \bar{a}\ln\frac{2B_m}{\bar{a}^{2}\left(R'_m\right)^2} + I_R(r_m) - \pi.
\end{align}
where
\begin{align}
R'_m&=\left\{\dfrac{\Theta ^2 (\ell-1) M \left[2 (\ell-1) M^2+2 M r_m-r_m^2\right]}{2 r_m^2 [2 (\ell-1) M+r_m]^2}+2 r_m\right\}\notag\\
&\quad\times\left\{2 \sqrt{\dfrac{\Theta ^2 \left[-2 (\ell-1) M^2+4 (\ell-1) M r_m+r_m^2\right]}{4 r_m [2 (\ell-1) M+r_m]}+r_m^2}\right\}^{-1}
\end{align}
The regular integral, which must be computed numerically for a given metric, is conveniently expressed in the original radial variable~\mbox{\cite{Igata:2025hpy}}:
\begin{align}
\label{eq:IRdef}
I_R(r_m) &= 2\int_{r_m}^{\infty}\left[
\frac{1}{\sqrt{-V_m(r)}} - \frac{\bar{a}\,R_m\,|R'(r)|}{R(r)\big(R(r)-R_m\big)}
\right]\mathrm{d}r\,,
\end{align}
where
\begin{align}
V_m(r) = \dfrac{D}{B}\left(1-\dfrac{f}{f_m}\right)
\end{align}
is the effective potential at $b = b_c$.  Both terms in the integrand diverge individually as $1/(r-r_m)$ for $r \to r_m$, but their difference is finite by construction of $\bar{a}$, so the integral converges.  Numerically, one starts the integration at $r_m + \varepsilon$ and verifies convergence as $\varepsilon \to 0$. This theoretical framework for strong-field deflection angles has been extensively applied in the literature; we therefore present only the essential steps here, referring the reader to~\cite{igata2025deflection,Igata:2025hpy,tsukamoto2023gravitational,034,37.6} for detailed derivations.

%%%%%%%%%%%%%%%%%%%%%%%%%%%%%%%%%%%%%%%%%%%%%%%%%%%%%%%%%%%%%%%%%%%%%%%%%%%%%%%%%%%%%%%%%%%%%%%%%%%%%%%%%%%%%%%%%%%%%%%%%%%%%%%%%%%%%%%%%%%%%%%%%%%%%%%%%%%%%%%%%%%%%%%%%%%%%%%%%%%%%%%%%%%%%%%%%%%%%%%%%%%%%%%%%%%%%%%%%%%%%%%%%%%%%%%%%%%%%%%%%%%%%%%%%%%%%%%%%%%%%%%%%%%%%%%%%%%%%%%%%%%%%%%%%%%%%%%%%%%%%%%%%%%%%%%%%%%%%%%%%%%%%%%%%%%%%%%%%%%%%%%%%%%%%%%%%%%%%%%%%%%%%%%%%%%%%%%%%%%%%%%%%%%%%%%%%%%%%%%%%%%%%%%%%%%%%%%%%%%%%%%%%%%%%%%%%%%%%%%%%%%%%%%%%%%%%%%%%%%%%%%%%%%%%%%%%%%%%%%%%%%%%%%%%%%%%%%%%%%%%%%%%%%%%%%%%%%%%%%%

\subsection{Numerical results and physical discussion}

We now apply the general formalism of the preceding subsection to the Kalb--Ramond metric~(\ref{metrictensorss}).  All quantities are computed by substituting the explicit functions $A(r)$, $B(r)$, $D(r)\big|_{\theta=\pi/2}$ from Sec.~\ref{Sec2} into Eqs.~\eqref{eq:bc}--\eqref{eq:IRdef} and performing numerical evaluation with the mass fixed at $M=1$.

\begin{table}[!tp]
\centering
\caption{{The strong deflection angle is computed for different values of the parameters $\Theta$ and $\ell$, with the black hole mass $M = 1$.}}
\begin{tabular}{|c|c|c|c|c|c|c|}
\hline
$\bar{b}$ & $\ell = 0.0$ & $\ell = 0.1$ & $\ell = 0.2$ & $\ell = 0.3$ & $\ell = 0.4$ & $\ell = 0.5$\\ \hline\hline
$\Theta = 0.0$ & -0.40023 & -0.54091 & -0.68964 & -0.84800 & -1.01814 & -1.20316 \\ \hline
$\Theta = 0.1$ & -0.39807 & -0.53789 & -0.68532 & -0.84160 & -1.00822 & -1.18687 \\ \hline
$\Theta = 0.2$ & -0.39164 & -0.52896 & -0.67263 & -0.82310 & -0.98042 & -1.14392 \\ \hline
$\Theta = 0.3$ & -0.38111 & -0.51446 & -0.65240 & -0.79448 & -0.93975 & -1.08746 \\ \hline
$\Theta = 0.4$ & -0.36673 & -0.49497 & -0.62585 & -0.75849 & -0.89230 & -1.02985 \\ \hline
$\Theta = 0.5$ & -0.34883 & -0.47117 & -0.59445 & -0.71810 & -0.84349 & -0.98768 \\ \hline
\end{tabular}
\label{Tab:bbar}
\end{table}

A thorough numerical analysis of the part $\bar{b}$ of the strong gravitational deflection angle for a fixed mass $M = 1$ is also given in Table \ref{Tab:bbar}, taking into account various non-commutativity parameter $\Theta$ and Lorentz-breaking parameter $\ell$ values. The findings clearly indicate a trend: $\bar{b}$ decreases as $\ell$ increases for a given $\Theta$. On the other hand, greater values of $\Theta$ cause the photon sphere radius to expand while $\ell$ remains constant. This trend is nearly the same as the outermost photon sphere radius's in Table \ref{Tab:rphoton}.

\begin{figure}[!tp]
\centering
\includegraphics[scale=0.8]{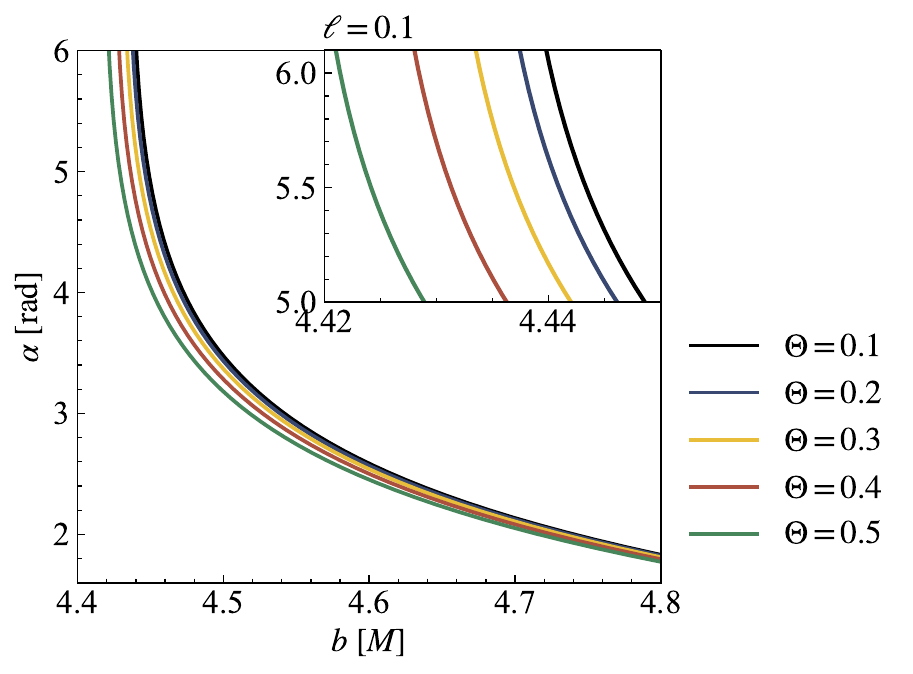}
\caption{The deflection angle $\alpha$ in the strong field limit is shown for different values of the parameter $\Theta$, when $\ell$ is fixed.}
\label{alpha_Theta}
\end{figure}

\begin{figure}[!tp]
\centering
\includegraphics[scale=0.8]{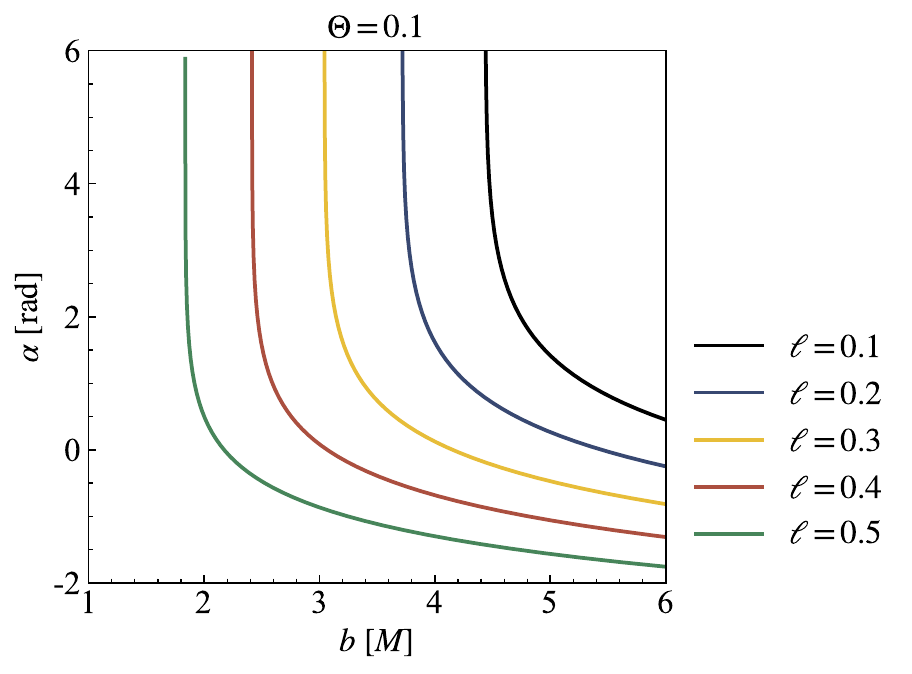}
\caption{The deflection angle $\alpha$ in the strong field limit is plotted for different values of the parameter $\ell$, when $\Theta$ is fixed.}
\label{alpha_ell}
\end{figure}

The influence of Kalb--Ramond gravity becomes evident in the deflection of light under strong--field conditions, as one could naturally expected. In Fig. \ref{alpha_Theta}, as the non--commutative parameter $\Theta$ increases from $0.1$ to $0.5$, the critical impact parameter shifts slightly to the left. For $b < b_{c}$, the deflection curves rise gradually, while for $b > b_{c}$, they shift downward. This behavior indicates that increasing $\Theta$ renders photon orbits unstable at smaller values of $b$, slightly enhancing the likelihood of capture. However, the shift remains small, appearing as a weak deformation of the classical trajectory.

In contrast, Fig. \ref{alpha_ell} reveals that the parameter $\ell$ has a much more pronounced effect. As $\ell$ increases from $0.1$ to $0.5$, the location of the divergence moves significantly to the right. Despite this, the overall trend in the deflection angle on either side of $b_c$ remains qualitatively similar to the case driven by $\Theta$.

The observed asymmetry in the effects of $\Theta$ and $\ell$ finds a clear physical interpretation within Igata's decomposition framework~\cite{Igata:2025hpy}. Since the Kalb--Ramond metric is strictly static (lacking any off-diagonal $g_{t\varphi}$ frame-dragging terms), the spacetime preserves time-reversal symmetry. Consequently, both the Weyl magnetic part $B_{(i)(j)}$ and the mixed Einstein component $G_{(0)(3)}$ vanish identically. The instability parameter $V''_{\mathrm{m}}$ is thus governed exclusively by the Weyl electric component $(E_{(2)(2)}-E_{(1)(1)})$--which encodes the local tidal shear--and the diagonal Einstein terms $(G_{(0)(0)}+G_{(3)(3)})$, representing the effective local energy density and transverse pressure.

To explicitly demonstrate the causal hierarchy of these deformation parameters, we evaluate the local geometric invariants at the equatorial photon sphere ($r=r_{\mathrm{m}}, \theta=\pi/2$). Expanding with respect to the noncommutative parameter $\Theta$, we obtain the leading-order analytical expressions evaluated at $r_{\mathrm{m}} \simeq 3M(1-\ell)$:
\begin{align}
E_{(2)(2)}^{\mathrm{m}} - E_{(1)(1)}^{\mathrm{m}} &= \frac{2-\ell}{18M^2(1-\ell)^3} + \mathcal{O}(\Theta^2), \label{eq:Weyl_E} \\
G_{(0)(0)}^{\mathrm{m}} &= -\frac{\ell}{9M^2(1-\ell)^3} + \mathcal{O}(\Theta^2), \label{eq:G00_m} \\
G_{(3)(3)}^{\mathrm{m}} &= 0 + \mathcal{O}(\Theta^2), \label{eq:G33_m} \\
V''_{\mathrm{m}} &= -\frac{2}{1-\ell} + \mathcal{O}(\Theta^2). \label{eq:Vpp_m}
\end{align}
These explicit asymptotic forms rigorously validate our physical interpretation. The Lorentz-violating parameter $\ell$ enters the expressions at linear order, acting as both a modifier to the background tidal shear~\eqref{eq:Weyl_E} and a non-trivial effective energy density~\eqref{eq:G00_m}. This fundamentally reshapes the underlying spacetime curvature and directly alters the baseline orbital instability from the Schwarzschild value ($-2$).

By contrast, the noncommutative parameter $\Theta$ enters all local geometric invariants strictly at $\mathcal{O}(\Theta^2)$. While $\Theta$ breaks the spherical symmetry of the background and induces additional axial anisotropy, manifesting through the higher-order corrections in the tidal shear $(E_{(2)(2)}^{\mathrm{m}} - E_{(1)(1)}^{\mathrm{m}})$, its influence is quadratically suppressed. This causal hierarchy---where $\ell$ linearly redefines the background curvature while $\Theta$ provides a suppressed quadratic geometrical deformation---rigorously explains why $\ell$ dominates the shifts in the optical observables $r_{\mathrm{m}}$, $b_c$, $\bar{a}$, and $\bar{b}$, whereas $\Theta$ remains a subleading perturbation.

%%%%%%%%%%%%%%%%%%%%%%%%%%%%%%%%%%%%%%%%%%%%%%%%%%%%%%%%%%%%%%%%%%%%%%%%%%%%%%%%

%%%%%%%%%%%%%%%%%%%%%%%%%%%%%%%%%%%%%%%%%%%%%%%%%%%%%%%%%%%%%%%%%%%%%%%%%%%%%%%%%%%%%%%%%%%%%%%%%%%%%%%%%%%%%%%%%%%%%%%%%%%%%%%%%%%%%%%%%%%%%%%%%%%%%%%%%%%%%%%%%%%%%%%%%%%%%%%%%%%%%%%%%%%%%%%%%%%%%%%%%%%%%%%%%%%%%%%%%%%%%%%%%%%%%%%%%%%%%%%%%%%%%%%%%%%%%%%%%%%%%%%%%%%%%%%%%%%%%%%%%%%%%%%%%%%%%%%%%%%%%%%%%%%%%%%%%%%%%%%%%%%%%%%%%%%%%%%%%%%%%%%%%%%%%%%%%%%%%%%%%%%%%%%%%%%%%%%%%%%%%%%%%%%%%%%%%%%%%%%%%%%%%%%%%%%%%%%%%%%%%%%%%%%%%%%%%%%%%%%%%%%%%%%%%%%%%%%%%%%%%%%%%%%%%%%%%%%%%%%%%%%%%%%%%%%%%%%%%%%%%%%%%%%%%%%%%%%%%%%%

\subsection{QNMs in eikonal limit}

In the eikonal approximation, the quasinormal mode spectrum is closely governed by the behavior of unstable circular photon orbits (UCPOs). The frequency's real component is dictated by the coordinate angular velocity $\Omega_{c}$, corresponding to either the prograde or retrograde trajectory. Meanwhile, the orbit's instability,  characterized by characterized by its Lyapunov exponent $\lambda$, defines the imaginary part of the frequency. As a result, one can formulate the QNM frequency as follows:
\ie
\omega_{QNM} = \Omega_{c} I - i\left(n+\frac{1}{2}\right)\lambda.
\fe
For large values of the multipole index $I$ (with $I \gg 1$) and a given overtone number $n$, the frequency of quasinormal modes can be parameterized accordingly. In this regime, the angular velocity $\Omega_{c}$ associated with the unstable circular photon orbit governs the oscillatory part of the spectrum. Explicitly, $\Omega_{c}$ takes the form:
\ie
\Omega_{c} = \frac{1}{b_{c}}.
\fe 
The correlation between the QNM parameters and the strong deflection limit coefficient is provided by
\begin{align}
\bar{a}_m = \frac{|\Omega_c|}{\lambda}.
\end{align}

A pronounced inverse relationship connects the quasinormal mode damping rate $\lambda$ and the photon trajectory divergence near the critical impact parameter $b_{c}$. When the non--commutative parameter $\Theta$ increases, $b_{c}$ experiences a modest displacement to smaller values. This minor shift induces comparable variations in both $\bar{a}_m$ and $\lambda$, signaling that the influence of non--commutativity acts as a weak perturbation on the spacetime geometry. In contrast, increasing the Lorentz--violating parameter $\ell$ substantially alters the potential well, producing a more dramatic leftward shift in $b_c$. This, in turn, simultaneously enhances $\bar{a}_m$ and suppresses $\lambda$, as evidenced in Figs. \ref{alpha_Theta} and \ref{alpha_ell}.

\section{\label{Sec13}Conclusion}

{

In this work, we focused on investigating the trajectory of light in connection with key optical features of a recently proposed black hole solution within Kalb--Ramond gravity, formulated in the framework of a non--commutative gauge theory of gravity \cite{heidari2025non}. Specifically, we examined the null geodesics, critical photon orbits, black hole shadows, gravitational lensing in both weak and strong field limits, and the properties of the topological photon sphere.

In particular, the behavior of null geodesics was examined to extract the fundamental features of the photon sphere and the resulting shadow structure. Initially, we numerically solved a system of partial differential equations to trace the trajectories of light rays, which were then plotted to visualize the photon paths. Subsequently, we computed the effective potential to identify the critical photon orbits. The analysis revealed the existence of a single photon sphere, given by
$r_{ph} \approx \, 3 M (1-\ell) + \frac{ \ell}{9 M}\Theta ^2,$
which aligned with the results obtained for the ``pure'' Kalb--Ramond black hole in Ref.~\cite{araujo2024exploring}. In general lines, it can be seen that increasing $\ell$ (for fixed $\Theta$) led to a decrease in the radius of the photon sphere, whereas increasing $\Theta$ (for fixed $\ell$) caused it to grow. Following this, we derived the expression for the shadow radius:
$R_{sh} \approx 3 \sqrt{3} M - \frac{\Theta^2}{8 \sqrt{3} M} + \left( -\frac{\Theta^2}{16 \sqrt{3} M} - \frac{9}{2} \sqrt{3} M \right) \ell,$
which exhibited the same qualitative behavior as the photon sphere. That is, $R_{sh}$ decreased with increasing $\ell$ and increased with larger values of $\Theta$. Also, The Event Horizon Telescope's observational data allowed researchers to place limits on $\Theta$ and $\ell$ by comparing them with expected shadow profiles.

The weak gravitational lensing was investigated via the Gauss--Bonnet theorem, relying fundamentally on the Gaussian curvature of the optical geometry. Additionally, the sign of the Gaussian curvature allowed us to verify the stability of the photon sphere, revealing that the critical orbit was unstable. For a fixed impact parameter, the deflection angle $\hat{\alpha}(b, \ell, \Theta)$ was found to increase. However, as the parameter $\ell$ increased, it exhibited a decreasing behavior. The analysis was extended to the strong deflection regime, where a logarithmic divergence emerged. As in the weak field case, the deflection angle $\alpha(b)$ increased with higher values of $\Theta$ (for certain ranges of $b$), while it decreased with increasing $\ell$. Moreover, the gravitational lensing predictions of observers were compared against the Event Horizon Telescope observational data from Sgr A*.

}

A promising direction within the non--commutative gauge framework involves deriving the charged extension of the Kalb--Ramond black hole solution \cite{duan2024electrically} and analyzing its associated gravitational properties, as explored in \cite{heidari2024impact,araujo2025antisymmetric}. Additionally, it would be valuable to construct new solutions in the context of bumblebee gravity \cite{Casana:2017jkc}, including its formulation within the metric--affine approach \cite{filho2023vacuum,amarilo2024gravitational,araujo2024exact}. These developments are currently being investigated by the present authors.

%%%%%%%%%%%%%%%%%%%%%%%%%%%%%%%%%%%%%%%%%%%%%%%%%%%%%%%%%%%%%%%%%%%%%%%%%%%%%%%%%%%%%%%%%%%%%%%%%%%%%%%%%%%%%%%%%%%%%%%%%%%%%%%%%%%%%%%%%%%%%%%%%%%%%%%%%%%%%%%%%%%%%%%%%%%%%%%%%%%%%%%%%%%%%%%%%%%%%%%%%%%%%%%%%%%%%%%%%%%%%%%%%%%%%%%%%%%%%%%%%%%%%%%%%%%%%%%%%%%%%%%%%%%%%%%%%%%%%%%%%%%%%%%%%%%%%%%%%%%%
\section*{Data availability statement}
There is no data available in this manuscript.
%%%%%%%%%%%%%%%%%%%%%%%%%%%%%%%%%%%%%%%%%%%%%%%%%%%%
\section*{Acknowledgments}
\hspace{0.5cm} A. A. Araújo Filho is supported by Conselho Nacional de Desenvolvimento Cient\'{\i}fico e Tecnol\'{o}gico (CNPq) and Fundação de Apoio à Pesquisa do Estado da Paraíba (FAPESQ), project No. 150891/2023-7. Partial funding for I. P. L. was provided by the National Council for Scientific and Technological Development (CNPq), under grant No. 312547/2023-4. I. P. L. acknowledges the networking support received through the COST Actions Bridging high and low energies in search of quantum gravity (BridgeQG, CA23130), Relativistic Quantum Information (RQI, CA23115) and Testing Fundamental Physics with Seismology (FuSe, CA24101), under the European Cooperation in Science and Technology (COST) framework. N. H. acknowledges the support of the COST Actions COSMIC WISPers in the Dark Universe: Theory, Astrophysics and Experiments (CosmicWISPers, CA21106), Addressing observational tensions in cosmology with systematics and fundamental physics (CosmoVerse, CA21136), and Bridging high and low energies in search of quantum gravity (BridgeQG, CA23130).

%%%%%%%%%%%%%%%%%%%%%%%%%%%%%%%%%%%%%%%%%%%%%%%%%%%%%%%%%%%%%%%%%%%%%%%%%%%%%%%%%%%%%%%%%%%%%%%%%%%%%%%%%%%%%%%%%%%%%%%%%%%%%%%%%%%%%%%%%%%%%%%%%%%%%%%%%%%%%%%%%%%%%%%%%%%%%%%%%%%%%%%%%%%%%%%%%%%%%%%%%%%%%%%%%%%%%%%%%%%%%%%%%%%%%%%%%%%%%%%%%%%%%%%%%%%%%%%%%%%%%%%%%%%%%%%%%%%%%%%%%%%%%%%%%%%%%%%%%%%%%%%%%%%%%%%%%%%%%%%%%%%%%%%%%%%%%%%%%%%%%%%%%%%%%%%%

	\bibliography{main}
	\bibliographystyle{unsrt}
	
\end{document}